\documentclass[a4paper,11pt, ]{article}
\pdfoutput=1

\usepackage{jcappub} 

\usepackage{subcaption}

\usepackage[T1]{fontenc} 
\usepackage{amsmath}
\DeclareMathOperator{\arcsec}{arcsec}

\usepackage[usenames, dvipsnames]{xcolor}
\usepackage{hyperref}
\usepackage[export]{adjustbox}
\usepackage{tikz}
\usetikzlibrary{shapes.geometric, arrows}

\usepackage{calc}      
\usepackage{setspace} 
\usepackage{fixltx2e}
\usetikzlibrary{shapes,arrows}
\usetikzlibrary{positioning}
\newcommand\norm[1]{\left\lVert#1\right\rVert}

\title{Forward Modeling of Spectroscopic Galaxy Surveys: Application to SDSS}

\author[a]{Martina Fagioli,}
\author[a]{Julian Riebartsch,}
\author[a]{Andrina Nicola,}
\author[a]{J{\"o}rg Herbel,}
\author[a]{Adam Amara,}
\author[a]{Alexandre Refregier,}
\author[b]{Chihway Chang,}
\author[a,c]{Laurenz Gamper,}
\author[a]{and Luca Tortorelli}

\affiliation[a]{Institute for Particle Physics and Astrophysics, ETH Z{\"u}rich, 8093 Z{\"u}rich, Switzerland}
\affiliation[b]{Kavli Institute for Cosmological Physics, University of Chicago, Chicago, IL 60637, USA}
\affiliation[c]{uSystems, Technoparkstrasse 2, 8406 Winterthur, Switzerland}

\emailAdd{martina.fagioli@phys.ethz.ch}

\abstract{Galaxy spectra are essential to probe the spatial distribution of galaxies in our Universe. To better interpret current and future spectroscopic galaxy redshift surveys, it is important to be able to simulate these data sets. We describe \texttt{U\textsc{spec}}, a forward modeling tool to generate galaxy spectra taking into account some intrinsic galaxy properties as well as instrumental responses of a given telescope. The model for the intrinsic properties of the galaxy population, i.e., the luminosity functions, and size and spectral coefficients distributions, was developed in an earlier work  for broad-band imaging surveys \citep{herbel2017}, and we now aim to test the model further using spectroscopic data. We apply \texttt{U\textsc{spec}} to the SDSS/CMASS sample of Luminous Red Galaxies (LRGs). We construct selection cuts that match those used to build this LRG sample, which we then apply to data and simulations in the same way. The resulting real and simulated average spectra show a good statistical agreement overall, with residual differences likely coming from a bluer galaxy population of the simulated sample. We also do not explore the impact of non-solar element ratios in our simulations. For a quantitative comparison, we perform Principal Component Analysis (PCA) of the sets of spectra. By comparing the PCs constructed from simulations and data, we find good agreement for all components. The distributions of the eigencoefficients also show an appreciable overlap. We are therefore able to properly simulate the LRG sample taking into account the SDSS/BOSS instrumental responses. The differences between the two samples can be ascribed to the intrinsic properties of the simulated galaxy population, which can be reduced by further improvements of our modelling method in the future. We discuss how these results can be useful for the forward modeling of upcoming large spectroscopic surveys.}

\keywords{Spectroscopic surveys, spectra simulations, principal components analysis}

\begin{document}

\maketitle
\flushbottom

\newpage

\section{Introduction}
\label{intro}

Although the cosmological principle ensures that the universe is homogeneous and isotropic when studied at sufficiently large scales, observations tell us that at smaller scales galaxies are not randomly and evenly distributed in space. Galaxies clump into clusters, and create voids, large areas of the universe which are empty, forming also complicated structures like filaments and sheets. This large scale structure depends both on the cosmology which describes the universe, and on galaxy properties. Three-dimensional maps that take into account the angular positions in the sky and the redshifts of galaxies are therefore a very powerful cosmological probe \citep{peebles1970, sunyaev1970, bond1984, coil2013}. This, combined with measurements of intrinsic galaxy properties such as colors, luminosities, morphologies, spectral types, or stellar masses, can also provide clues about galaxy formation and evolution \citep{madgwick2003, zehavi2005, zehavi2011}.

A large sample of tracers of such large scale structures is needed to extract information about the galaxy clustering pattern and its relation with galaxy properties \citep{tegmark1997, goldberg1998, eisenstein1998, hong2012}. Ongoing and upcoming wide-field galaxy surveys such as the Dark Energy Survey\footnote{\url{https://www.darkenergysurvey.org}.} (DES), the Kilo-Degree Survey\footnote{\url{http://kids.strw.leidenuniv.nl}.} (KiDS) and the survey of the Large Synoptic Survey Telescope\footnote{\url{https://www.lsst.org}.} (LSST) provide a wealth of photometric data. However, the errors associated with photometric redshift measurements make spectroscopic redshift surveys also necessary. Galaxy redshift surveys such as the Deep Extragalactic Evolutionary Probe (DEEP2) \citep{weiner2005}, the Very Large Telescope Deep Survey (VVDS) \citep{garilli2008} and the Baryon Oscillation Spectroscopic Survey (BOSS) \citep{schlegel2009}  within the Sloan Digital Sky Survey (SDSS) III \citep{eisenstein2011} already provide measurements of the galaxy clustering power spectrum.

In light of upcoming large spectroscopic redshift surveys, such as the extended BOSS (eBOSS\footnote{\url{http://www.sdss.org/surveys/eboss/}}) \citep{dawson2016, zhao2016}, the Dark Energy Spectroscopic Instrument (DESI\footnote{\url{http://desi.lbl.gov/}}) \citep{desi1, desi2}, the Wide Field Infrared Survey Telescope (WFIRST\footnote{\url{https://wfirst.gsfc.nasa.gov}}), and ESA's Euclid satellite\footnote{\url{http://www.euclid-ec.org/}} \citep{laureijs2011}, it is necessary to be able to interpret the increasingly precise and numerous data from such cosmological surveys and to forecast the science performances of the experiments, for example, understanding redshift fitting routines or the reliability of photometric redshift estimates. Simulations will play a key role in this scenario. Spectroscopic surveys such as SDSS/BOSS, the 6df Galaxy Survey (6dFGS), the Big Baryon Oscillation Spectroscopic Survey (BigBOSS) and 4m Multi-Object Spectroscopic Telescope (4MOST) \citep{dejong2012} have developed simulation tools to prepare their observing strategies and improve their data reduction performances (see e.g., \citep{campbell2004, blanton2003, schlegel2011, boller2012}). \citep{nord2016} developed the SPectrOscopic KEn Simulation (SPOKES), an end-to-end simulation facility for spectroscopic cosmological surveys. 

In this paper, we describe \texttt{U\textsc{spec}} and its performances in simulating realistic galaxy spectra. \texttt{U\textsc{spec}} takes as inputs a galaxy model, described in details in \citep{herbel2017}, and the instrumental setup of a given telescope, and outputs redshifted, noisy galaxy spectra. This allows us to forward model a spectroscopic galaxy sample, and to compare it to an existing one after having performed the same cuts on both. The forward modeling approach is becoming a widely used technique \citep{refregier2014, herbel2017, bruderer2017}, and consists of generating observable quantities from an astrophysical model containing, for instance, the evolution of the luminosity functions of red and blue galaxies with cosmic time (see e.g. \citep{herbel2017}). A similar approach has also been performed recently for image simulations with narrow-band photometric data from the Physics of the Accelerating Universe Survey, (PAUS\footnote{\url{https://www.pausurvey.org}}), as shown in \citep{tortorelli2018}.

Here, we choose to simulate a sample of Luminous Red Galaxies (LRGs) from the SDSS surveys. LRGs are a widely used tracer of large scale structure \citep{eisenstein2005, hutsi2006, padmanabhan2007, almeida2008}, and their spectra can be well approximated by a linear combination of templates and coefficients \citep{norberg2002, cappellari2004, cappellari2017}. Even though LRGs are considered easier to model than blue star-forming galaxies, still they present some challenges. For example, it has been already shown in 2003 by \citep{eisenstein2003} that describing SDSS LRGs with templates constructed with solar chemical abundances of some atomic species is not sufficient to fully model their spectra. Indeed, massive LRGs are found to have values of $\alpha$- to Fe-elements abundance ratios ($[\alpha$/Fe]) enhanced with respect to solar values ($[\alpha$/Fe] > 0), both at $z=0$ \citep{thomas2003, lee2005, thomas2005, thomas2010}, and at higher redshift \citep{onodera2015, lonoce2015, fagioli2016, kriek2016}. However, modeling the features of individual galaxies is beyond the scope of our paper. Our purpose is to simulate a population of LRGs whose average properties are comparable to a real SDSS red galaxy population. The goal is to include such spectra simulations into a framework which includes broad-band and narrow-band data coming from different photometric and spectroscopic surveys, in order to better constrain the galaxy redshift distribution $n(z)$ presented in \citep{herbel2017}, which relied on broad-band photometry only. The redshift distribution of galaxies of different types is of fundamental importance to tackle cosmological probes such as galaxy clustering and cosmic shear, as shown for example in \citep{nicola2016, nicola2017}.

We compare our simulated sample to a real one from SDSS, after applying the appropriate cuts to both. The resulting stacked galaxy spectrum is compared to a stacked galaxy spectrum coming from the SDSS/BOSS survey. We also perform a Principal Component Analysis which shows the agreement between the simulations and the data, showing that the method is able to reproduce the main properties of the LRGs we aimed to study, with the limitation of being able to probe a narrower range of individual galaxy properties, as expected from a simplified galaxy model as the one used here. 

The paper is structured as follows. In Section~\ref{model}, we describe the model from which the basic ingredients to simulate galaxies are drawn. In Section~\ref{data}, we describe the data set and the reasons behind the cuts applied to the data. In Section~\ref{fulluspec}, we describe \texttt{U\textsc{spec}} and the instrumental setup added to the model galaxies in order to generate realistic spectra. In Section~\ref{results}, we compare our simulated data to real galaxy spectra, both through comparing stacked real and simulated spectra and through a Principal Component Analysis of the two populations. In Section~\ref{conclusions}, we present our conclusions.
\\
\\
Throughout this work, we use a standard $\Lambda$CDM cosmology with $\Omega_m$ = 0.3, $\Omega_\Lambda$ = 0.7 and $H_0 = 70$ km s$^{-1}$ Mpc$^{-1}$.

\section{Galaxy population model}\label{model}
To simulate galaxy spectra, we need basic galaxy properties as inputs. The model from which such properties are drawn is fully described in \citep{herbel2017}, and further clarified in \citep{tortorelli2018}. In this section, we review the aspects of the model necessary for simulating galaxy spectra given an input galaxy population. For a more complete description of the model we refer the reader to \citep{herbel2017}. 

\subsection{Galaxy luminosity functions}

The galaxies are drawn from galaxy luminosity functions $\phi$ (see e.g. \citep{johnston2011, beare2015}). We refer to a galaxy luminosity function as the number of galaxies $N$ for comoving volume $V$ and absolute magnitude $M$:

\begin{equation}
\phi(z,M)=\frac{dN}{dM\,dV}
\end{equation}

where $z$ denotes redshift. The functional form of $\phi$ is taken to be a Schechter function \citep{schechter1976}, {i.e.},

\begin{equation}
\phi (\mathrm{z,M}) = \frac{2}{5} \log{(10)} \phi_* 10^{\frac{2}{5} (M_* - M) (\alpha + 1)} \exp{\left[ -10^{\frac{2}{5} (M_* - M)} \right]}
\end{equation}

Here, the parameters $M_*$ and $\phi_*$ evolve as: 

\begin{equation}
\begin{split}
M_*(z) &=M_{*, \mathrm{slope}}\ z + M_{*, \mathrm{intcpt} } \\
\mathrm{\phi_*(z)} &= \mathrm{\phi_{*,amp} \exp{(\phi_{*,exp}\ z)}} \\
\end{split}
\end{equation}

where $M_{*, \mathrm{slope}}$ and $ M_{*, \mathrm{intcpt} }$ are the slope and the intercept that set the evolution of $M_{*}$, while $\mathrm{\phi_{*,exp}}$ and $\mathrm{\phi_{*,amp}}$ are the exponential decay rate and the amplitude that determine the evolution of ${\phi_*}$. All these values are listed in Table~\ref{tab: lumfunc}.

The model parameters (or functioning point, as defined in \citep{herbel2017} and reported in Appendix B in \citep{tortorelli2018}) are listed in Table~\ref{tab: lumfunc}. Blue and red galaxies are drawn from separate and evolving luminosity functions. This distinction is done through their Specific Star Formation Rates (SSFRs, SSFRs $ > 0.007$ yr$^{-1}$ for blue, SSFRs $ < 0.001$ yr$^{-1}$ for red galaxies). The SSFRs used to initially construct the model are taken from the New York University Value-Added Galaxy Catalog (NYU-VAGC\footnote{\url{http://sdss.physics.nyu.edu/vagc/}}, \citep{blanton2005}), as described in detail in \citep{herbel2017}. Note that this separation into blue and red galaxies is not used in the selection cuts described below to define our galaxy sample, as our cuts are based on colors and magnitudes (and redshifts). The redshifts and absolute magnitudes are obtained by sampling from the corresponding luminosity function. The appropriate sets of coefficients used to construct galaxy spectra are chosen taking into account such properties (see following section).

\begin{table}
\centering
\begin{tabular}{c c c}
 & {\textbf{Luminosity Functions}} & \\
\hline
\hline
& {Red} & {Blue} \\
\hline

$\alpha$ & -0.5 & -1.3 \\
\hline
$M_{*,slope}$ & -0.70798041 & -0.9408582 \\
\hline
$M_{*,intcpt}$ & -20.37196157 & -20.40492365 \\
\hline
$\mathrm{\phi_{*,exp}}$ & -0.70596888 & -0.10268436 \\
\hline
$\mathrm{\phi_{*,amp}}$ & 0.0035097 & 0.00370253 \\

 & {\textbf{Dirichlet Parameters}} & \\
\hline
\hline
& {Red} & {Blue} \\
\hline
$\mathrm{\alpha_1}$ & 1.62158197 & 1.9946549 \\
\hline
$\mathrm{\alpha_2}$ & 1.62137391 & 1.99469164\\
\hline
$\mathrm{\alpha_3}$ & 1.62175061 & 1.99461187\\
\hline
$\mathrm{\alpha_4}$ & 1.62159144 & 1.9946589\\
\hline
$\mathrm{\alpha_5}$ & 1.62165971 & 1.99463069\\
\hline
\end{tabular}
\caption{Model parameters used in this work, as also reported in Table 3 in \citep{tortorelli2018}.}
\label{tab: lumfunc}
\end{table}

\subsection{Spectral energy distributions}\label{ktemp}

The next step is to model the Spectral Energy Distributions (SEDs). We model the SEDs of galaxies as linear combinations of templates $T_{i}(\lambda)$ weighted by coefficients $c_i$, where $T_{i}(\lambda)$ are suitably chosen templates and $c_i$ come from a Dirichlet distribution \citep{dirichlet} of order five. The choice of a Dirichlet distribution is motivated by the construction of \citep{herbel2017}'s model. In brief, the parameters $c_i$ are drawn by a Dirichlet distribution and used to construct the spectrum. The spectrum itself is then scaled in order to match its corresponding absolute magnitude $M$, drawn from the appropriate luminosity function. As $c_i$ are used to scale the spectrum considering its absolute magnitude, their ratios are relevant, while their overall sum can be arbitrarly chosen. \citep{herbel2017} then chose to restrict their sum always to 1, which the is basic property of a Dirichlet distribution. The Dirichlet distribution is parametrized by five model parameters which we refer to as $\alpha_i$. These are different for red and blue galaxies and evolve with redshift $z$ as: 

\begin{equation}
\alpha_i(z)=(\alpha_{i,0})^{1-z/z_1}\times({\alpha_{i,1}})^{z/z_1}
\end{equation}

where $\alpha_{i,0}$ is chosen to sample the galaxy population at redshift $z=0$, and  $\alpha_{i,1}$ at redshift $z = z_1 > 0$.

The templates used are the five $kcorrect$ templates presented in \citep{blanton2007} and shown in the rightmost column of Figure 4 in their paper. The $kcorrect$ templates are based on the Bruzual $\&$ Charlot stellar evolution synthesis models \citep{bruzual2003} and computed though Nonnegative Matrix Factorization (NMF, \citep{lee99}). 485 spectral templates are used to construct the final five global templates we use in this analysis. 450 are Single Stellar Population (SSP) from Bruzual $\&$ Charlot (2003) with Chabrier Initial Mass Function (IMF) \citep{chabrier2003} , with metallicity grid $Z=$ 0.0001, 0.0004, 0.004, 0.008, 0.02, and 0.05, and 25 ages between 1 Myr and 13.75 Gyrs. The remaining 35 templates are from are from MAPPINGS-III \citep{kewley2001} models of emission from ionized gas. In the end, we compute the SEDs as 

\begin{equation}
\mathrm{SED}=\sum_{i=1}^5 c_i T_{i}
\end{equation}

where $T_i$ are the final 5 $kcorrect$ templates and $c_i$ come from a Dirichlet distribution as described above.

\subsection{Catalog generator}
The basic galaxy population ingredients (coefficients $c_i$, redshifts and magnitudes) described above are generated and stored as described in \citep{herbel2017}. In \citep{herbel2017}, the galaxy catalogs were used in order to simulate astronomical images, with the \texttt{Ultra Fast Image Generator} (\texttt{U\textsc{fig}}) \citep{berge2013, bruderer2016, bonnett2016, leistedt2016, herbel2017, bruderer2017}. Galaxy catalogs can be generated modifying the filters to compute the output magnitudes, and the extinction given by Milky Way dust can be added. This is the first time these catalogs are used to generate galaxy spectra.

\section{Data: SDSS/BOSS}\label{data}

As a starting point, we choose a red galaxy sample that is widely used for large scale structure studies and whose properties are expected to be easier to model. We thus select LRGs from Data Release 13 (DR13) of the SDSS/BOSS survey \citep{albareti2017}. The imaging and spectroscopic data of this survey were obtained at the 2.5m telescope of the Apache Point Observatory (APO) in Sunspot, New Mexico \citep{gunn2006}, with respectively a wide field mosaic CCD camera \citep{gunn1998} and a twin multi-object fiber spectrograph. The BOSS survey (Baryon Oscillation Spectroscopic Survey) \citep{schlegel2009}  within the SDSS III \citep{eisenstein2011} uses an upgrade of the SDSS spectrograph. We describe its relevant features in the sections below.

\subsection{SDSS photometry}\label{sdssphot}

We make use of the \texttt{Composite Model Magnitudes (cModelMag)}\footnote{\url{https://www.sdss.org/dr12/algorithms/magnitudes/}} of the SDSS catalog \citep{stoughton2002}. The \texttt{cModelMag} magnitudes are a good total flux indicator for each band. These are the most suitable choice of magnitude to be compared to our simulated magnitudes, as the latter are not aperture matched in different bands. In the following analysis, we do not consider the \textit{u}-band, as the \textit{u}-band magnitude measurements have large photometric errors for SDSS red galaxies (see e.g. \citep{blanton2007}). Note that we do not correct the magnitudes used here for the reddening caused by Galactic dust, but we include this effect in our modeling of the LRGs.

\subsection{SDSS spectroscopy}

The spectra used in this analysis are taken from the BOSS survey. Here we highlight the relevant features of the spectra we employ; for a full description of the characteristics of these data we refer the reader to \citep{smee2013}. 

In the BOSS survey, the number of fibers per plate has been increased to 500 (from the previous 320 of SDSS) so that a total of 1000 objects is observed per exposure. Of these fibers, 895 are dedicated to science targets, 100 to sky and standard stars, and 5 to repeated targets. The BOSS fiber size is $2\arcsec$. This is the most suitable choice for high redshift galaxies (as in this case, up to $z=0.7$) in order to maximize the signal-to-noise while keeping the sky background contamination low. We use spectroscopic redshifts from \citep{bolton2012}. The wavelength range extends from 3,650 to 10,400 \AA{}. However, part of the blue wavelengths ($\sim200$\AA{}) are excluded from the analysis below due to fringing. The same applies to the reddest wavelengths of the spectrum ($\sim2500$\AA{}) due to the prominent sky background residuals. For a more detailed description of noise effects, such as read-out and shot noise, and effects of flux loss such as atmospheric transmission, see Section~\ref{instrumentalresp}. For a more detailed description of the sky background and resolving power effects, see Section~\ref{skymodel}.

\subsection{Sample selection}\label{sampsel}

Here, we describe how we selected the final sample of the galaxies that we analyzed. We emulate cuts as in the SDSS/BOSS CMASS\footnote{\url{http://www.sdss3.org/dr9/algorithms/boss_galaxy_ts.php}} sample \citep{dawson2013}. CMASS, which stands for Constant Mass, aims at selecting luminous massive galaxies which are homogeneously distributed in stellar mass at $z > 0.4$, as described in \citep{maraston2013}. This homogeneity in stellar mass is obtained by adopting the \citep{maraston2009} model of passively evolving galaxies of different stellar masses as a function of cosmic time \citep{maraston2013}. Color cuts are applied in the  $(g-r)$ and $(r-i)$ plane in order to isolate high redshift galaxies, in the approximate redshift range $0.4<z<0.7$. The color cuts explicitly applied in SDSS-I/II Cut-II \citep{eisenstein2001} and 2SLAQ \citep{Cannon2006} to select red galaxies are not applied in CMASS. This is the reason why in our final sample galaxies with sign of gas emission (like e.g., $\mathrm{[O II]}$ and $\mathrm{[O III]}$ emission lines) are included. However, galaxies with visible emission lines only account for about 4$\%$ of the total CMASS sample (see e.g., \citep{thomas2013} and \citep{reid2016}).
\\
\\
First, we exclude galaxies with $g-,\,r-\,$ or $z-$band magnitudes above the SDSS magnitude limit in those bands, namely:

\begin{itemize}
\item $g<22.2$
\item $r<22.2$
\item$z<20.5$
\end{itemize}
Cuts on the $i-$band magnitude are already included in the final CMASS color/magnitude cuts applied below. No cuts are applied in the $u-$band (see Section~\ref{sdssphot}).
\\
\\
In order to emulate the CMASS sample, the following cuts in magnitude, color and redshift are applied:

\begin{itemize}
\item $i_{fib2}<21.5$
\item $17.5<i<19.9$
\item $r-i<2$
\item $d_\perp>0.55$
\item $i<19.86+1.6 \times (d_\perp-0.8)$
\item $0.4<z<0.7$
\end{itemize}
where $d_\perp = (r-i) - (g-r) / 8$, which is the distance perpendicular to the locus of the galaxy colors in the $r-i$ vs. $g-r$ color plane. This ensures the exclusion of low redshift galaxies  from the sample. The cuts in the $i-$band define the faint and bright limits. $i_{fib2}$ is the measurement of the flux contained within the aperture of a spectroscopic fiber in $i-$band ($2\arcsec$ in the case of the BOSS spectrograph). 

A sharp cut in redshift ($0.4<z<0.7$) is also introduced by us in order to ensure a closer match between the redshift distribution of simulated and real galaxies (see Section~\ref{redshifts} below). This cut is introduced to focus on the performances of the spectra simulations. In our future work, we will rely only on photometric quantities, as spectroscopic properties such as redshifts need to be measured on spectra themselves, which are the data products we seek to simulate.

\subsection{Catalogs comparison}\label{compinput}

In this section, we present the comparison between the simulated and real galaxy catalogs. The simulated properties are derived from the model described in Section~\ref{model} and are given as input in order to simulate galaxy spectra. Here we compare those simulated input properties and the data. In Section~\ref{sampsel} we listed the cuts in magnitude and redshift spaces applied to SDSS galaxies to mimic the CMASS sample. The same cuts have been applied to the simulated galaxies with the appropriate modifications (see below).

\subsubsection{Magnitudes}\label{magnitude}

\begin{figure}
\centering

	\includegraphics[width=120mm]{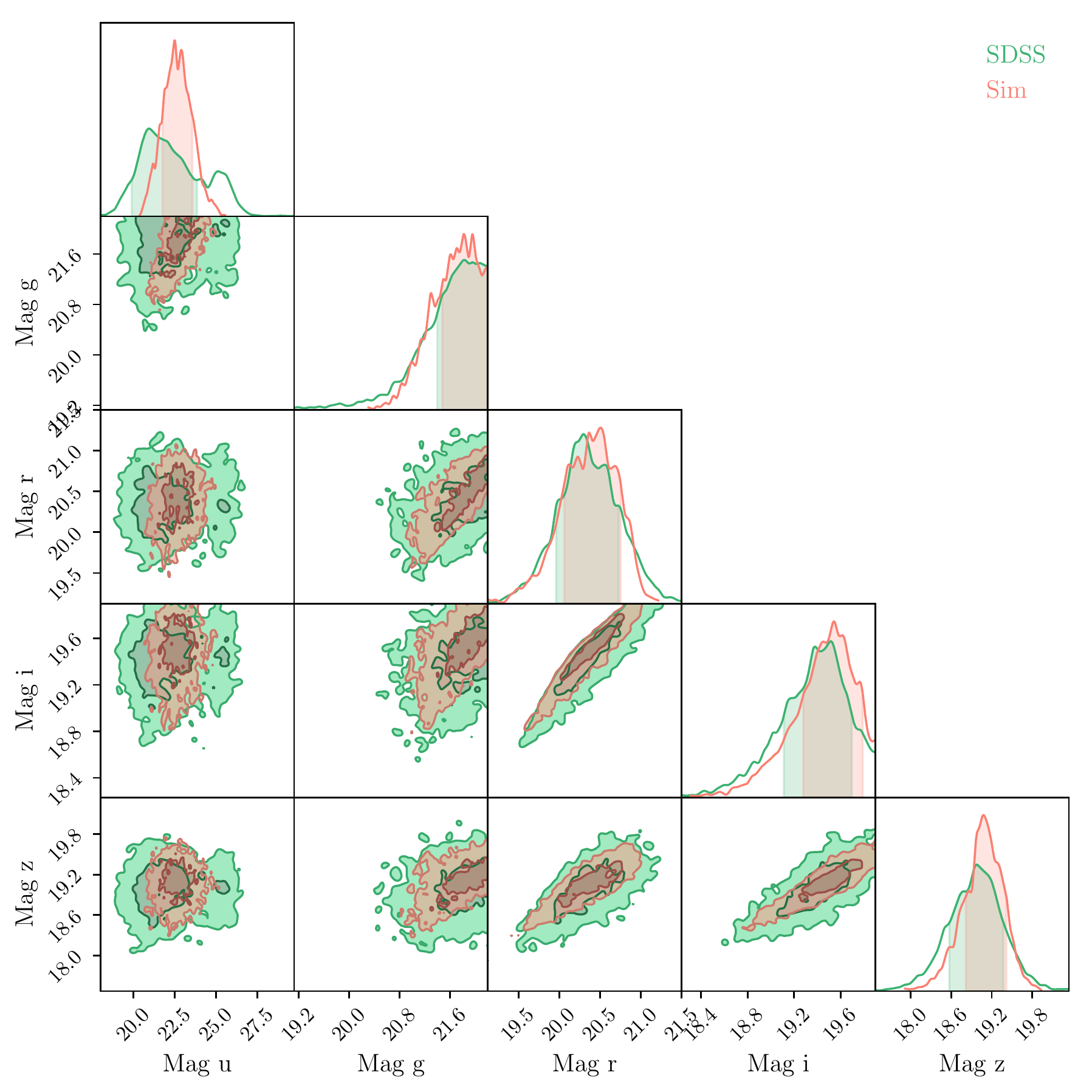}

    \caption{SDSS (green) and simulated (red) magnitude distributions. The simulated magnitudes have been derived using SDSS filters and the errors added considering the SDSS magnitude error distributions.}
    \label{fig: color_color}
\end{figure}

The magnitudes here employed for the simulated galaxies have been computed using SDSS filters. We correct these `intrinsic' galaxy magnitudes by fitting a linear relation between them and \texttt{MAG AUTO} magnitudes computed with \texttt{S\textsc{ource}} \texttt{E\textsc{xtractor}} (\texttt{SE\textsc{xtractor}}) \citep{bertin1996} on SDSS-like simulated images \citep{bruderer2016, herbel2017}. These magnitudes are noise free. To add realistic uncertainties to our magnitudes, we take those from the real SDSS magnitudes we employ. We find a correlation between SDSS magnitudes and their associated uncertainties, which we model with a linear relationship. As expected, fainter magnitudes have larger uncertainties. We construct Gaussians centered in the fitted value of uncertainty in each magnitude bin, having as standard deviation the scatter around the fitted values. We randomly draw uncertainties from these distributions. These uncertainties are then added to our noise-free simulated magnitudes.

The data magnitudes we employ are affected by reddening caused by dust. Therefore, we also add this effect on simulated magnitudes using reddening maps from \citep{schlegel1998}. With regard to the cut presented in Section~\ref{sampsel} on the \texttt{fiber2Mag} magnitude, the \texttt{fiber2Mag} is assumed to be the same as the $i-$band magnitude  for the simulated galaxies.

Figure~\ref{fig: color_color} shows the comparison between real (green) and simulated (red) magnitudes in all SDSS bands  ($ugriz$). It can be seen from Figure~\ref{fig: color_color} that the distributions of the real and simulated galaxy population magnitudes are similarly centered. In all bands, the SDSS magnitude distributions occupy a larger parameter space. The $u-$band shows a difference in the cut-off of the distributions at faint magnitudes. This is mostly due to a population of galaxies which exhibit $u-$band magnitudes above the magnitude limit of SDSS ($u<22$). We did not exclude these galaxies from the analysis for the known issues in the SDSS $u-$band outlined in~\ref{sdssphot}. The wider variety of SDSS magnitudes could be explained for example by more various stellar populations which are likely to be present in the sample of real galaxies, with the possible addition of a a metal-poor old stellar component as shown in \citep{maraston2009}; \citep{tojeiro2012} indeed found that CMASS galaxies have a wide variety of intrinsic galaxy properties. \citep{wake2006} also found differences in real and simulated colors in the 2SLAQ LRGs sample, where their higher redshift sample shows bluer $g-r$ colors than the used models. Also, massive galaxies are found to be enhanced in $\alpha$-elements with respect to the solar values, also increasing the variety of properties of real galaxies with respect to the simulated ones. Finally, the noise modeling could concur in generating such a difference.

\subsubsection{Redshifts}\label{redshifts}

Redshifts are drawn from luminosity functions and assigned to simulated galaxies during the galaxy catalog generation, as described in Section~\ref{model} and \citep{herbel2017,tortorelli2018}. Figure~\ref{fig: redshifts} shows the comparison between the real (green) and the simulated (red) spectroscopic redshift distributions. The simulated redshift distribution is shifted towards lower redshifts, with a median redshift of 0.524 for SDSS galaxies and median redshifts of 0.506 for simulated galaxies {(i.e., $\Delta z = 0.018$). In future work, the parameters of the input model can be adjusted to improve the match of the two distributions.

In this first analysis of \texttt{U\textsc{spec}}, we force a match between the redshift distributions of the real and the simulated samples. The matching is done such that in every bin of redshift there is the same number of objects for both the real and the simulated galaxies, randomly chosen from a parent sample. However, it is worth noting that the difference in $n(z)$ is related to the parameters controlling the redshift evolution of the luminosity functions. For a more detailed discussion on this aspect, see Appendix~\ref{appendix}. After this matching, the total number of galaxies to be analyzed for both the real and the simulated sample is 2433 objects.

\begin{figure}
\centering

	\includegraphics[width=88mm]{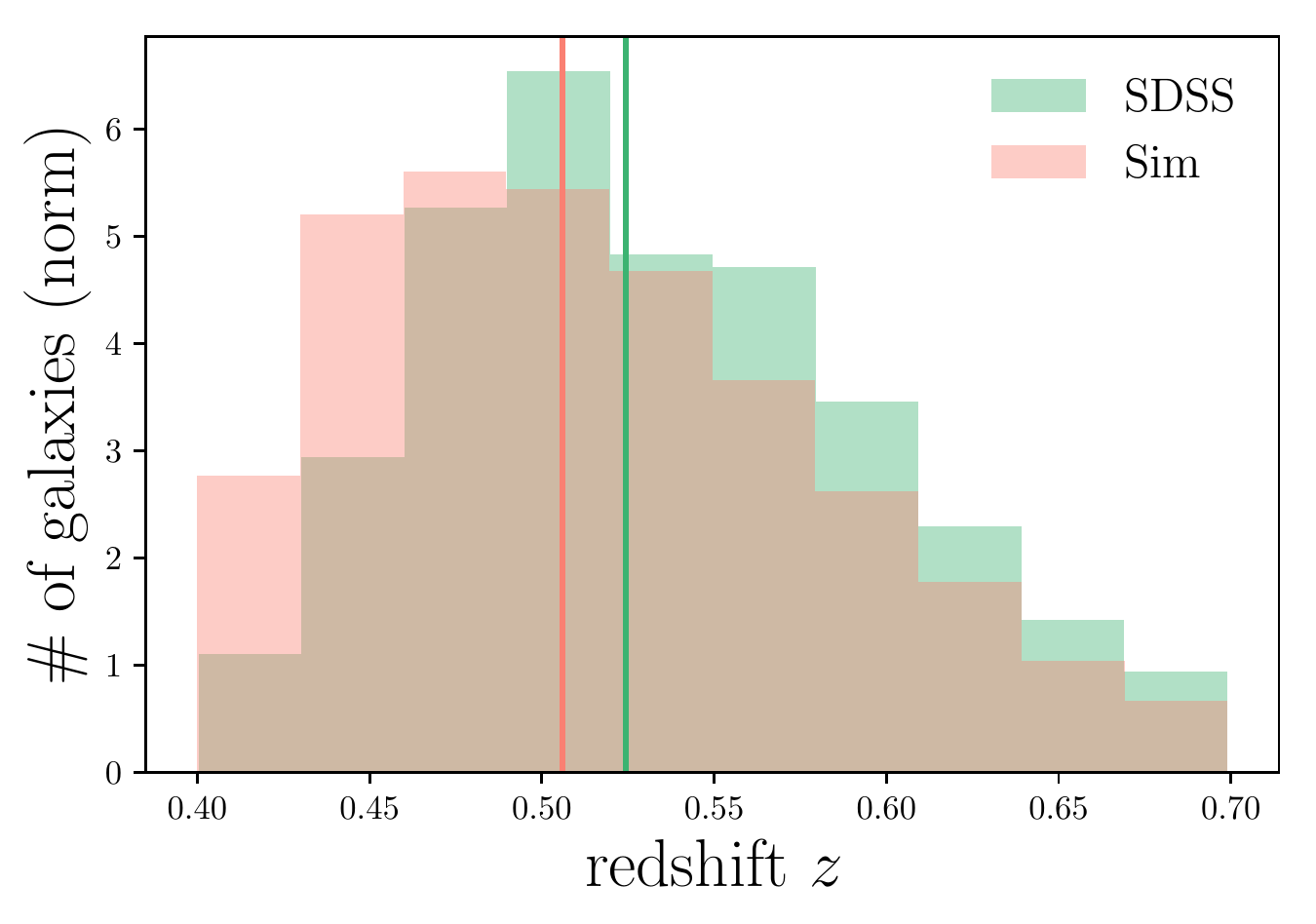}

    \caption{Spectroscopic redshift distributions for SDSS (green) and simulations (red). The vertical lines show the medians of the two distributions. }
    \label{fig: redshifts}
\end{figure}

\section{Spectra simulations: \texttt{U\textsc{spec}}}\label{fulluspec}

\texttt{U\textsc{spec}} simulates galaxy spectra of an experiment given its location, instrument and a cosmological model. All the steps taken in order to construct galaxy spectra are illustrated in the flowchart in Figure~\ref{fig: flowchart}. 

First, the spectrum ${f}(\lambda)$ of a galaxy is constructed as a linear combination

\begin{equation}
{f}(\lambda) = \sum_{i=1}^5 c_{i}T_i(\lambda) 
\end{equation}
where $T_i(\lambda)$ are the 5 \textit{kcorrect} templates from \citep{blanton2007}, and the coefficients $c_{i}$ are described in Section~\ref{ktemp}. This produces a noise-free, rest-frame galaxy model spectrum. The spectrum is then shifted to the observed frame given the redshift $z$ provided by the simulated galaxy catalog. The dimming effect due to shift in wavelengths, i.e., the fact that the flux enclosed in a wavelength bin $\Delta\lambda$ must be assigned to a broader wavelength bin $\Delta\lambda(1+z)$, and the dimming due to the luminosity distance $d_L$ of the sources, are applied \citep{appenzeller2009}.

\subsection{Instrumental response}\label{instrumentalresp}

Noise is added to the constructed galaxy model spectra. The instrumental effects which are included into the simulated SDSS-like spectra are listed below:

\begin{itemize}

\item \textit{Read-out noise}: A read-out noise of $3e^{-}$/pixel is assumed (see as a reference \citep{smee2013} for read-out noise in the BOSS spectrograph). The read-out noise is computed per unit wavelength, taking into account the instrumental resolution, and a random normal realization of it is added to the model galaxy and sky spectra in photons (see the flowchart in Figure~\ref{fig: flowchart}). 

\item \textit{Shot noise}: The shot noise is the poisson random realization of the model galaxy or sky spectrum. Specifically, a poisson realization of the $f_{gal}+f_{sky}$, where $f$ stands  in this case for flux expressed in photons, is created. A different random poisson realization of $f_{sky}$ is then subtracted, in order to simulate a realistic sky background subtraction, as sky from a different fiber in the same plate is subtracted from the galaxy in the SDSS survey. We do not account for differences in the sky background given by different locations in the SDSS plate, which we assume to be negligible. For a detailed description of the sky model, see Section~\ref{skymodel}.

\item \textit{Transmission curve}: The transmission loss due to atmosphere and instrument is taken into account. We have used the \texttt{`spthroughput'} routine of the \texttt{`idlspec2D'} spectroscopic reduction pipeline\footnote{\url{http://www.sdss3.org/dr8/software/products.php}} built by Princeton University and flux calibration files to create the throughput of the instrument. The galaxy and sky model spectra are first multiplied for the atmospheric transmission. The final, sky-subtracted spectrum is then divided by the atmospheric transmission as to mimic the steps of the data reduction (see flowchart in Figure~\ref{fig: flowchart}).
\end{itemize}

\tikzstyle{decision} = [diamond, draw, fill=blue!20, 
    text width=6.5em, text badly centered, node distance=3cm, inner sep=0pt]
    
\tikzstyle{block} = [rectangle, draw, 
    text width=7em, text centered, rounded corners, minimum height=1.5em]
    
\tikzstyle{blockmed} = [rectangle, draw, 
    text width=20em, text centered, rounded corners, minimum height=1.5em]
\tikzstyle{mycircle} = [circle, thick, draw=orange, minimum height=4mm]

\tikzstyle{line} = [draw, -latex']
\tikzstyle{cloud} = [draw, ellipse,fill=red!20, node distance=3cm,
    minimum height=2em]

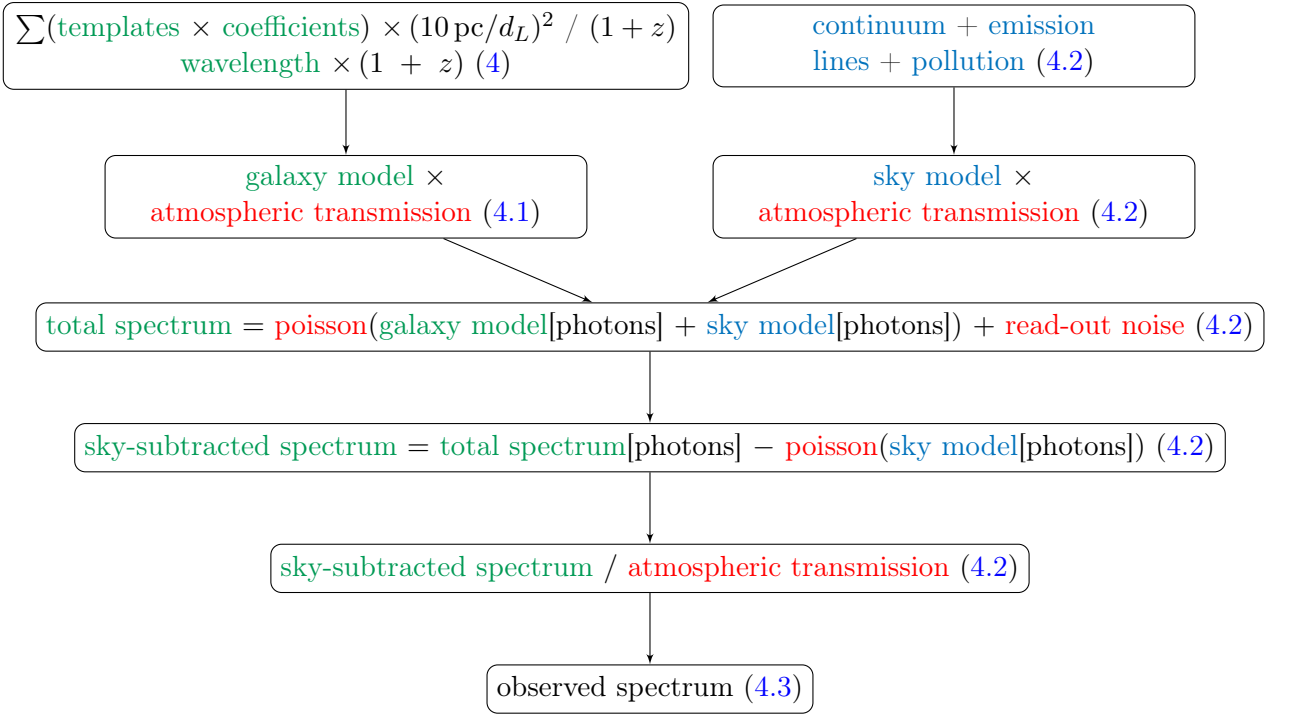
\begin{figure}
\centering
\begin{tikzpicture}[align=left, node distance = 2cm, auto]

    \node [block, text width=22.9em] (temp) {$\sum$(\textcolor{ForestGreen}{templates} $\times$ \textcolor{ForestGreen}{coefficients}) $\times\,(10\,\mathrm{pc} / d_{L})^2$ / $(1+z)$ \\
    \textcolor{ForestGreen}{wavelength} $\times\,(1+z)$ (\ref{fulluspec})};

    \node [block, right of=temp, node distance=8cm, text width=15.9em] (sky) {\textcolor{RoyalBlue}{continuum} + \textcolor{RoyalBlue}{emission lines} + \textcolor{RoyalBlue}{pollution} (\ref{skymodel})};

    \node [blockmed, below  of=temp, text width=15.9em, node distance=2.0cm] (specatm)
             {\textcolor{ForestGreen}{galaxy model} $\times$\\ \textcolor{red}{atmospheric transmission} (\ref{instrumentalresp})};
             
    \node [blockmed, below  of=sky, text width=15.9em, node distance=2.0cm] (skyatm)
             {\textcolor{RoyalBlue}{sky model} $\times$ \\ \textcolor{red}{atmospheric transmission} (\ref{skymodel})};  
             
 \path (specatm) -- node (poisson1) [rectangle, draw,  rounded corners, below=1.4cm]{\textcolor{ForestGreen}{total spectrum} $=$ \textcolor{red}{poisson}(\textcolor{ForestGreen}{galaxy model}[photons] $+$ \textcolor{RoyalBlue}{sky model}[photons]) $+$ \textcolor{red}{read-out noise} (\ref{skymodel})}(skyatm);
 \path (specatm) -- node (poisson2) [rectangle, draw,  rounded corners, below=3.0cm]{\textcolor{ForestGreen}{sky-subtracted spectrum} $=$ \textcolor{ForestGreen}{total spectrum}[photons] $-$ \textcolor{red}{poisson}(\textcolor{RoyalBlue}{sky model}[photons]) (\ref{skymodel})}(skyatm);
 \path (specatm) -- node (atmfinal) [rectangle, draw,  rounded corners, below=4.6cm]{\textcolor{ForestGreen}{sky-subtracted spectrum} $/$ \textcolor{red}{atmospheric transmission} (\ref{skymodel})}(skyatm);
 \path (specatm) -- node (final) [rectangle, draw,  rounded corners, below=6.2cm] {{observed spectrum} (\ref{finalspec})}(skyatm);

    \path [line] (temp) -- (specatm);
    \path [line] (sky) -- (skyatm);
    \path [line] (specatm) -- (poisson1);
    \path [line] (skyatm) -- (poisson1);
    \path [line] (poisson1) -- (poisson2);
    \path [line] (poisson2) -- (atmfinal);
    \path [line] (atmfinal) -- (final);

\end{tikzpicture}
\caption{Flowchart illustrating the construction of the final galaxy model spectra. All instrumental effects (read-out and shot noise, and effects of atmospheric transmission) are indicated in red. In light blue, all steps relative to the construction and use of the sky model spectrum are shown. All steps involving the use of the galaxy model only are shown in green. Poisson$(\cdot)$ indicates a random poisson realization of the flux given in photons (shot noise).}
\label{fig: flowchart}
\end{figure}

\subsection{Sky model}\label{skymodel}

\begin{figure}
\centering

	\includegraphics[width=120mm]{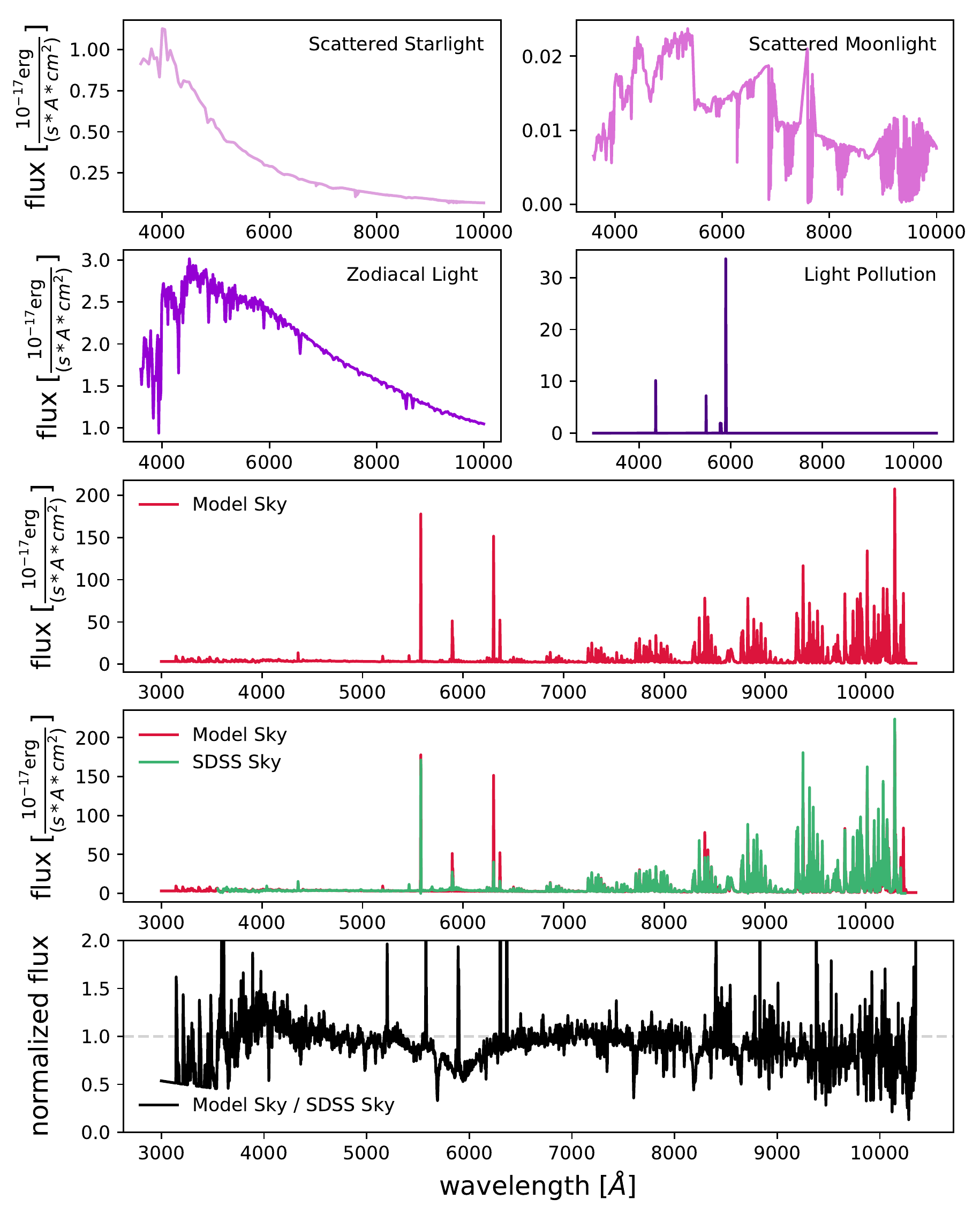}

    \caption{Sky model spectrum. The first four panels show the contributions to the continuum (from scattered starlight, scattered moonlight in dark time, and zodiacal light) and emission lines coming from night pollution which are not visible in Paranal spectra. The model sky emission lines come from the UVES Atlas of Paranal sky lines. The total (noise-free) model sky is shown in red. The comparison with a random (noisy) sky taken from the SDSS sample is shown in the second last panel. The residuals between the two are shown in the last panel. The bump around 6000 \AA{} is due to atmospheric pollution due Na and Ne lamps \citep{law2016}, which are not present in Paranal sky spectra.}
    \label{fig: skymodel}
\end{figure}

The emission lines of the night sky are important contaminants for astronomical observations. It is therefore important to properly model the sky spectrum in order to estimate its effective impact on the noise associated to the galaxies simulated by \texttt{U\textsc{spec}}. 

The night sky spectrum is recorded in every single astronomical observation in SDSS. Although its variability makes it difficult to model if one wants to simulate the precise intensity of the features of the night sky at a given moment of the night and position in the sky, the central wavelengths of the emission lines are constant in time and space, and can therefore be easily modeled. For the purpose of forward modeling the sky, a sky spectrum at a given moment of time or position in space is not needed. The poisson realization of a night sky background which includes the common sky lines and continuum is sufficient for  purpose of mimicking a realistic sky subtraction. Below we list the input parameters for our simulated sky model; in Figure~\ref{fig: skymodel} we show our (noise-free) model sky spectrum model, its individual components and the comparison with a real (noisy) random BOSS sky spectrum. 

\begin{itemize}
\item{{Line Spread Function (LSF): }} We compute the LSF (i.e., the broadening due to the instrumental resolving power $\mathrm{R} = \lambda / \Delta\lambda$ as a function of wavelength $\lambda$) on randomly observed sky spectra in the BOSS survey. We fit Gaussians to a series of evenly distributed sky lines along the wavelength direction. We then derive a linear relation between the FWHM of those lines and their central wavelengths. The relation we derive is in agreement with the resolving power provided by BOSS ($\mathrm{R}\sim1500$ in the blue range, $\mathrm{R}\sim2500$ in the red range \citep{smee2013}). We use this relation to determine the width of the sky lines in our model spectrum.
\newpage
\item{{Emission Lines:}}
\begin{itemize}
\item \textit{UVES Atlas of Paranal sky lines\footnote{\url{https://www.eso.org/observing/dfo/quality/UVES/pipeline/sky_spectrum.html}}}: The sky lines central wavelengths and intensities are taken from \citep{hanuschik2003}. The atlas of lines in the optical and near-IR wavelength range has been acquired by UVES, the echelle spectrograph at the 8.2-m UT2 telescope of the Very Large Telescope (VLT). {The absolute intensities of sky line emission depend on the time and location of the observations. Here we use the UVES line intensities as a reference in order to construct a realistic sky model spectrum. They are not meant to precisely match a particular sky spectrum observed in SDSS. This is the reason why a disagreement in some of the emission lines intensities is visible in Figure~\ref{fig: skymodel} between the real and the simulated sky spectrum.} 

\item \textit{Light pollution emission lines}: \citep{osterbrock1996} list emission lines tracing light pollution, such as HgI 5461, 5770, 5791 \AA{} and components of the NaI 5890, 5896 \AA{} lines indicative of both high-pressure and low-pressure Sodium lamps. These lines are not included in the UVES atlas of sky lines as in a dark site like Paranal there is no trace of such elements in the atmosphere. For this reason, we fit Gaussians to these sky lines from a real SDSS spectrum, deriving their peak intensities. We then construct a mock spectrum with using these sky lines only, using the LSF described above and the intensities as described here, and add it to the total spectrum.

\end{itemize}

\item{{Continuum:}} The continuum for our model sky has been computed with the \texttt{SkyCal Web Application}\footnote{\url{https://www.eso.org/observing/etc/bin/gen/form?INS.MODE=swspectr+INS.NAME=SKYCALC}} from the ESO Sky Calculator \citep{noll2012, jones2013}. This includes:

\begin{itemize}

\item{\textit{Scattered Moonlight}: Scattered moonlight has a stronger effect on the continuum if the observing date is close to full Moon. This is not the case for the observations of CMASS, which have been taken during dark time. However, as particularly the blue wavelengths are affected by it, it can not be neglected.}

\item{\textit{Zodiacal Light}: Zodiacal light is coming from interplanetary dust grains scattering sunlight. Here we choose values for ecliptic latitude $\beta$ and heliocentric ecliptic longitude $\lambda-\lambda_\odot$ for targets at the zenith, with airmass = 1. A strong continuum coming from zodiacal light would be expected for low absolute values of such coordinates \citep{noll2012}. }

\item{\textit{Scattered Starlight}:  Starlight is scattered in the atmosphere. The distribution of stars reaches a peak when it gets close to the centre of the Milky Way. Therefore, the scattering model required for this kind of distribution is that for extended sources. This component of the continuum is minor compared to the other two main components mentioned above. As a consequence, computing a mean continuum spectrum it is sufficient for an exposure time calculator application, such as the one used here \citep{noll2012}.}

\end{itemize}
\end{itemize}
A demonstration of the impact of the sky model on the data analysis is given in Appendix~\ref{appendixb}.

\subsection{Construction of the final \texttt{U\textsc{spec}} spectrum}\label{finalspec}
As described in the flowchart in Figure~\ref{fig: flowchart} and in the previous sections, a model sky-subtracted galaxy spectrum is built starting by a linear combination of 5 $kcorrect$ templates and coefficients assigned to the galaxies, as described in Section~\ref{model}. The simulated spectrum includes read-out and shot noise and the effects of the atmospheric transmission. At this point, the magnitude in the $r-$band is computed and compared to the input $r-$band. A warning is generated in case the difference between the two exceeds $5\%$. This happens for $\sim0.002\%$ of the generated galaxy spectra, which are however included in the final sample. The final fluxes, input magnitudes, redshifts and output $r-$band magnitudes are stored in order to be compared with those from real data.

\section{Results}\label{results}

\subsection{Stacked spectra comparison}\label{stacked}

\begin{figure}
\centering

	\includegraphics{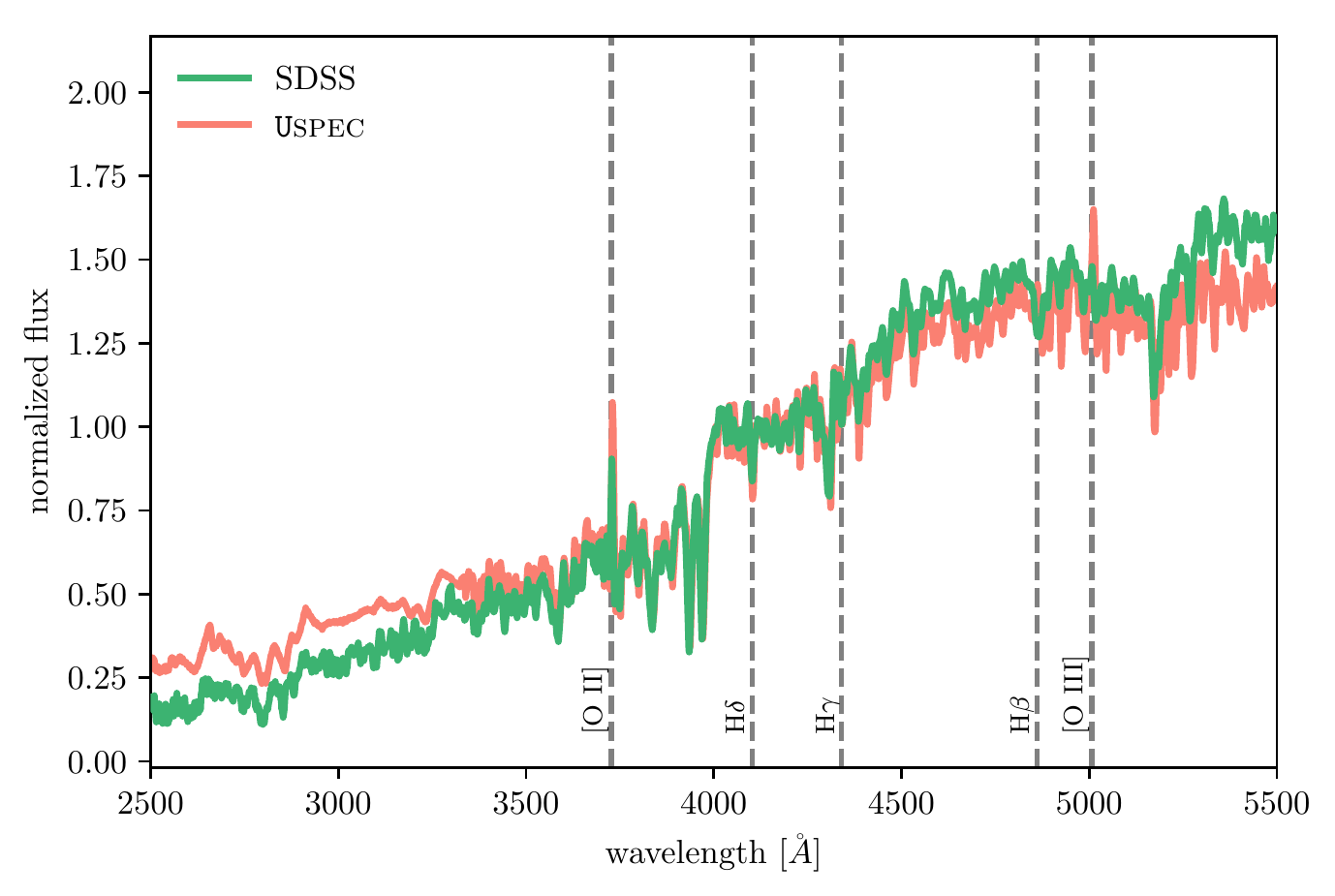}

    \caption{{Average stacked LRGs spectra for both SDSS (green) and \texttt{U\textsc{spec}} (red) samples. Vertical dashed lines indicate the position of strong features, such as [O II] and [O III] emission lines and the Balmer absorption lines.}}
        \label{fig: spectra_comp}

\end{figure}

In order to compare real and simulated galaxy spectra, we {construct mean (composite) spectra as in \citep{onodera2012, onodera2015, fagioli2016}}. A total of 2433 galaxy spectra is used for both the SDSS and the \texttt{U\textsc{spec}} samples. {The composite spectra are computed by shifting each individual spectrum to the rest-frame. Each spectrum is then normalized by the mean flux at $4,100< \lambda < 4,700$ \AA{} wavelengths. In this region, the spectra of red galaxies are flat and no strong features are expected. The spectra are then interpolated onto a 1 \AA{} linearly spaced wavelength grid. In each $\Delta \lambda$ bin, the mean flux among all the spectra is then taken. These steps are followed in the construction of both the real and the simulated stacked spectra. For SDSS}, the spectra are corrected for Galactic extinction following the extinction curve for diffuse gas from \citep{odonnell1994} with $R_v=3.1$, and using the Galactic $E(B-V)$ values from the maps of \citep{schlegel1998}. This is needed as simulated galaxy spectra do not include Galactic dust. We do not correct for any internal dust extinction, since passive galaxies are expected to have negligible intrinsic dust. Figure~\ref{fig: spectra_comp} shows the comparison between the two spectra. The average spectra for both galaxy samples are clearly those of red galaxies with an old stellar population, revealing features such as the Ca II H $\&$ K lines, the G-band at 4300 \AA{}, the Balmer absorption lines and the break at $4,000$ \AA{}. The Mg lines are also clearly visible in both spectra. 

Overall, the agreement between the two average spectra is reasonable as the goal of this paper is to show the capabilities of \texttt{U\textsc{spec}} to simulate realistic galaxy spectra given the inputs for cosmological purposes. Nonetheless, it is visible that the \texttt{U\textsc{spec}} average spectrum shows a somewhat flatter shape and a stronger emission in [O II] and [O III], an indication of an overall bluer population than the SDSS sample. This difference in the overall population can be explained by looking at the difference in the color-color space between the two populations. Figure~\ref{fig: color_comp} shows the comparison between SDSS and \texttt{U\textsc{spec}} galaxies in the $g-i$ vs. $g-r$ color-color space. The Figure shows that the centroids of the two distributions have a small offset. The SDSS color-color distribution clearly spans a wider parameter space. This is probably due to the wider variety of intrinsic galaxy properties (see also \citep{tojeiro2012}) of the real galaxy population; for example, different chemical properties as reported in \citep{thomas2003, maraston2009}, especially in the [$\alpha$/Fe] abundances, which are found to be enhanced with respect to solar values in massive galaxies as those studied here, both at $z=0$ \citep{thomas2003, lee2005, thomas2005, thomas2010}, and at higher redshift \citep{onodera2015, lonoce2015, fagioli2016, kriek2016}. Similar differences have been also reported in \citep{eisenstein2003, thomas2003, wake2006, maraston2009, tojeiro2012}, as outlined in detail in the text above. See also Appendix~\ref{appendixc} for a comparison of [$\alpha$/Fe]-related spectral features in the real and simulated samples. This explains the difference in the overall shape of the stacked spectra and the stronger emission in [O II] for the simulated galaxies with respect to the SDSS ones, and also shows the ability of \citep{herbel2017} and \texttt{U\textsc{spec}} to generate realistic spectra given some input properties. This difference in the spectra can be used as a diagnostic to fix the differences in the input galaxy properties in \citep{herbel2017}. This can be achieved in future work with an ABC (Approximate Bayesian Computation) optimization of the model parameters, as discussed in Section~\ref{conclusions}. These differences, however, highlight the need of further quantifying the agreement of the two populations studied, which we now turn to.

\begin{figure}
\centering

	\includegraphics[width=98mm]{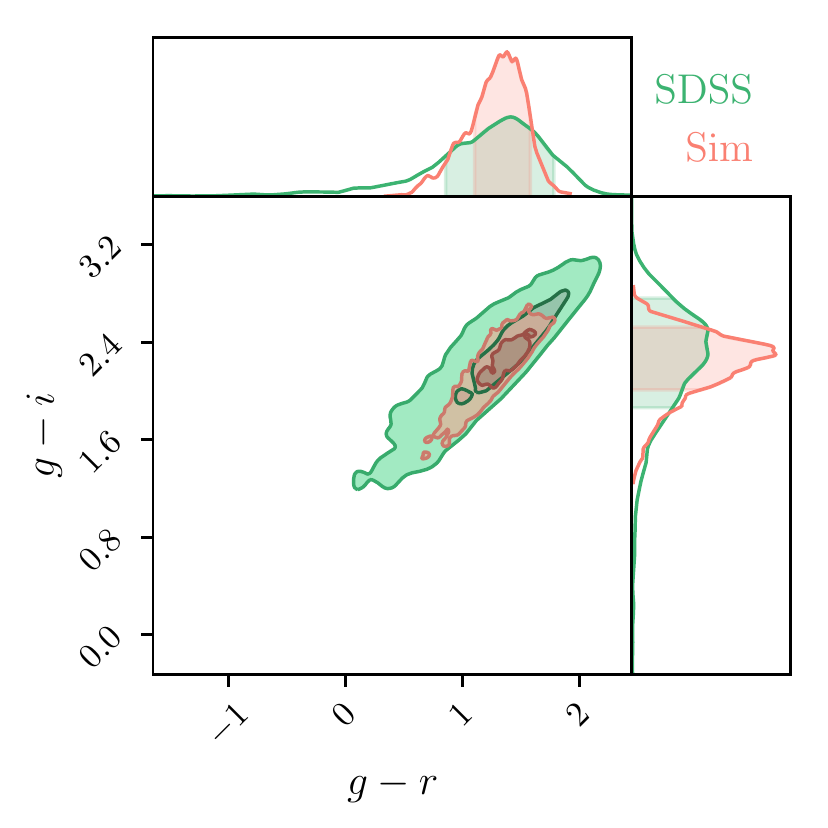}

    \caption{Figure showing the $g-i$ vs. $g-r$ color-color diagram, showing the comparison between SDSS and simulated galaxies. See Appendix~\ref{appendixc} for a comparison of detailed spectral features in the real and simulated samples.}

    \label{fig: color_comp}
\end{figure}

\subsection{Principal Component Analysis (PCA)}
As discussed above, we need to quantify the differences between the two populations of real and simulated galaxies. Galaxy spectra contain large amount of information, as each galaxy spectrum is described by 3469 data points. A useful approach to this problem is therefore trying to reduce its dimensionality.

The Karhunen-Lo\'eve transform, also commonly called Principal Component Analysis (PCA), is a technique which is widely used to reduce the dimensionality of big data sets \citep{pcareview}. Its application in astronomy has been exploited in details (see e.g., \citep{efstathiou1984, murtagh1987, yip2004, chen2012}). Applying PCA to spectroscopy basically consists in representing the spectra as a lower dimensional set of eigenspectra \citep{connolly1995}. The eigenspectra are obtained by finding a matrix $U$ such that 

\begin{equation}
\label{eqinit}
U^{T}CU=\Lambda
\end{equation}
where $\Lambda$ is  the diagonal matrix containing the eigenvalues of the correlation matrix $C$ constructed from the spectra. No weights are included in this analysis. We solve this problem through Singular Value Decomposition (SVD). In both real and simulated spectra, we mask the regions where strong sky lines are expected, namely at 5578.5 \AA{}, 5894.6 \AA{}, 6301.7 \AA{}, 6364.5 \AA{}, 7246 \AA{}, with a FWHM of 15 \AA{} for each sky line. See Appendix~\ref{appendixb} for a discussion on how PCA performs without masking those lines. The first 200 \AA{} in the blue and about 2500 \AA{} in the infrared regions of the spectra have also been excluded from the analysis as they are severely dominated by residual sky features. The same analysis is applied to both the SDSS spectral sample and the \texttt{U\textsc{spec}} simulated spectra. Each galaxy spectrum $f(\lambda)$ can be constructed as follows:

\begin{equation}
\label{specdata}
{f}(\lambda)=\sum_j \mathrm{a}_{j}\phi_j(\lambda)\,~~\mathrm{for \,data}
\end{equation}

\begin{equation}
\label{specsim}
{f'}(\lambda)=\sum_j \mathrm{b}_{j}\psi_j (\lambda)\,~~\mathrm{for \,simulations}
\end{equation}
where a$_{j}$ (b$_{j}$) are the expansion coefficients, or eigencoefficients (see Section~\ref{coeff} below) and $\phi_j (\lambda)$ ($\psi_j (\lambda)$) are the eigenspectra for the data (simulations).

The first five principal components comparison for both samples is shown in Figure~\ref{fig: pc_comparison}. The grey areas show the masked regions where strong sky lines are expected. Overplotted are the $g-,\,r-\,$ and $i-$bands from SDSS. The PCA conducted independently for the data and the simulations shows good agreement in {all} components. The choice of using five components is motivated by the initial construction of the \texttt{U\textsc{spec}} spectra which are built from the five $kcorrect$ templates (Section~\ref{fulluspec}). As it can be seen in the Figure, the first two components capture most of the physical information coming from the spectra. Even if individual features are not visible, as the spectra are analyzed in the observed frame, the overall shape of red galaxies spectra is clearly visible in the first two PCA components. The three higher PCA components all show similar patterns, for both the SDSS and the \texttt{U\textsc{spec}} sample. In those, the characteristic bump at $\sim6100$ \AA{}, which is the position of the 4000 \AA{} break shifted by the median redshifts of the two populations, is clearly visible, together with other redshifted features such as [O II] emission, the G-band and the H$\beta$ absorption, all broadened due to the variety of the redshifts of the two samples. 

{PCA has already been applied to the study of the CMASS sample in \citep{chen2012}. \citep{chen2012} decompose each individual observed and model spectrum into their principal components representations. In other words, they compute their eigencoefficients (see below) and compare their distributions. These projections are then related to a variety of galaxy properties. Figure 2 in their work shows the first 7 principal components. It is interesting to note how they also find the first component to trace the overall shape of the spectrum, while higher order components more closely trace individual features. In their work, individual features are clearly seen as the whole analysis in conducted in rest-frame. Our observed frame analysis makes it more difficult to see individual features, as they are mostly averaged out by the variety of redshifts in the sample. In this work, no attempt is done to recover physical properties of the galaxy population, as it is beyond the scope of the paper. Our main purpose is to decompose both the real and the simulated galaxy populations into two distinct and independent basis sets. In the following section, we describe a method to compare these two independent basis sets and quantify their agreement.} 

\begin{figure}
\centering

	\includegraphics[width=100mm]{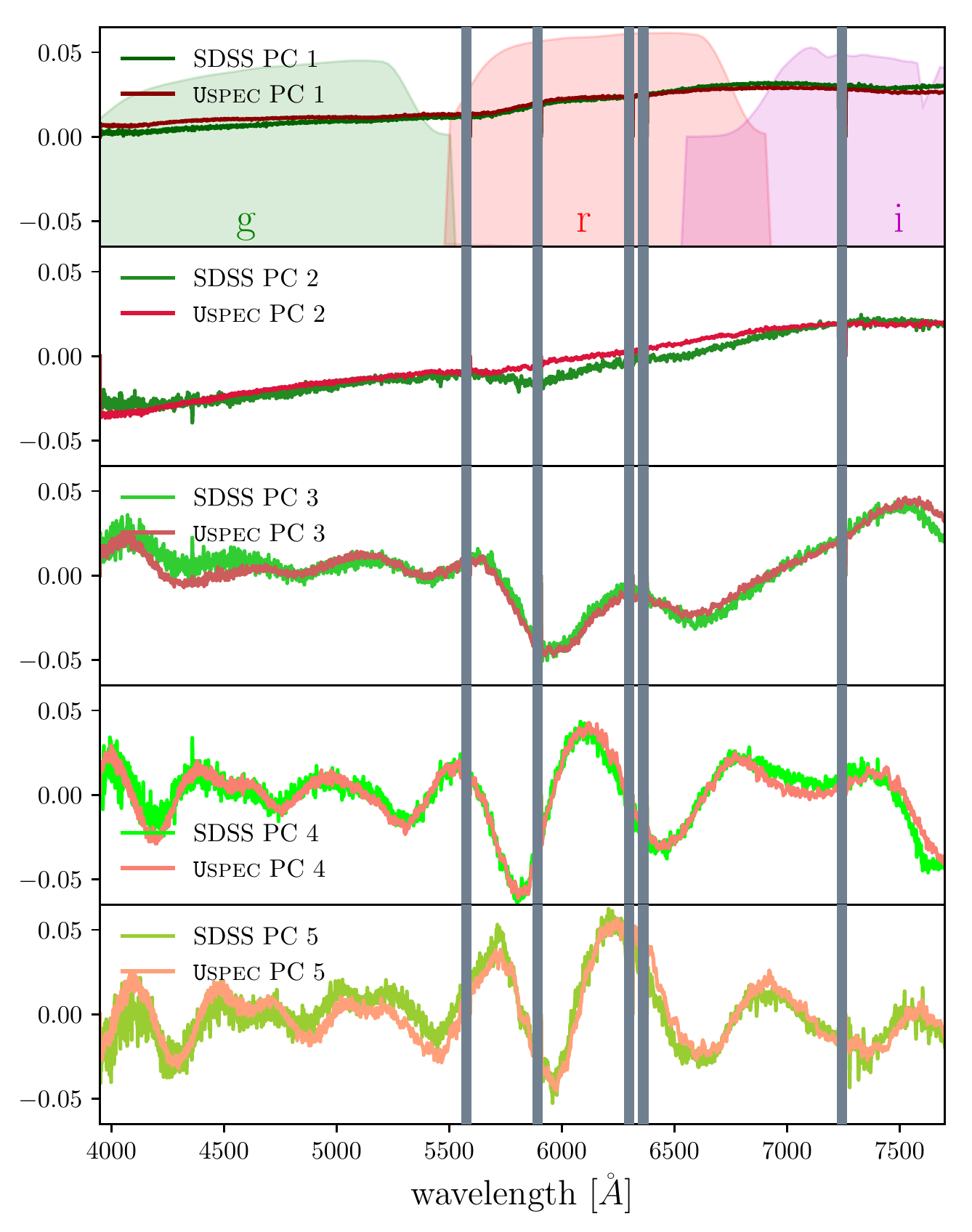}

    \caption{Principal components comparison between SDSS and \texttt{U\textsc{spec}} galaxies. The first component captures most of the signal coming from the spectra. The grey areas show the masked regions where strong sky lines are expected. The SDSS filter curves for $g-,\,r-,\,i-$ are overplotted.}
    \label{fig: pc_comparison}
\end{figure}

\subsubsection{Mixing matrix}
\label{mixingmatrix}
By definition, solving the eigenvalue problem of Equation~\ref{eqinit} means that the basis set  $\phi_i (\lambda)$ and $\psi_i (\lambda)$ are two sets of orthonormal basis. This means that we can define a mixing matrix M$_{ij}$ such that

\begin{equation}
\mathrm{M}_{ij}=\int\phi_i(\lambda)\psi_j(\lambda)~d\lambda
\end{equation}
so that, if $\phi_i (\lambda)\equiv\psi_i (\lambda)$, $\mathrm{M}_{ij}$ reduces to

\begin{equation}
\mathrm{M}_{ij}=\delta_{ij}
\end{equation}

In other words, in the ideal case, i.e., if real and simulated data were described by the same basis set, the mixing matrix would be the identity matrix. Figure~\ref{fig: pc_matrix} shows a graphical representation of the mixing matrix between the real and the simulated spectra presented above. In the Figure, the numbers in the boxes show the scalar products between the different components. As expected, the first components show better agreement than higher order components. The non diagonal elements of the matrix are significantly smaller than the diagonal ones up to the {fifth} component. A distance metric can be defined in order to assess how far the mixing matrix is from being diagonal. We choose to use the ratio between the product of the diagonal elements of matrix M$_{ij}$ and its determinant, $\varepsilon$. For a diagonal matrix, such a ratio should be one. In our case

\begin{equation}
\varepsilon=\norm {\frac{\Pi(\mathrm{M}_{ii})}{\det(\mathrm{M}_{ij})} }= 0.95
\end{equation}
which shows how similar the two basis sets independently used to describe real and simulated spectra are. In Appendix~\ref{appendix}, we describe how the mixing matrix changes when we do not match the redshift distributions of the two populations. In particular, it is worth noting how $\varepsilon$ varies for a difference in the median redshift between the two populations of $\Delta z=0.018$, going from {0.95 to 0.77}. In Appendix~\ref{appendixb}, we show how $\varepsilon$ changes with not masking the strong sky emission lines, dominating all principal components except for the first one. The effect of sky lines should be taken into account as they might strongly influence the distance metrics $\varepsilon$ introduced here.

\begin{figure}
\centering

	\includegraphics[width=85mm]{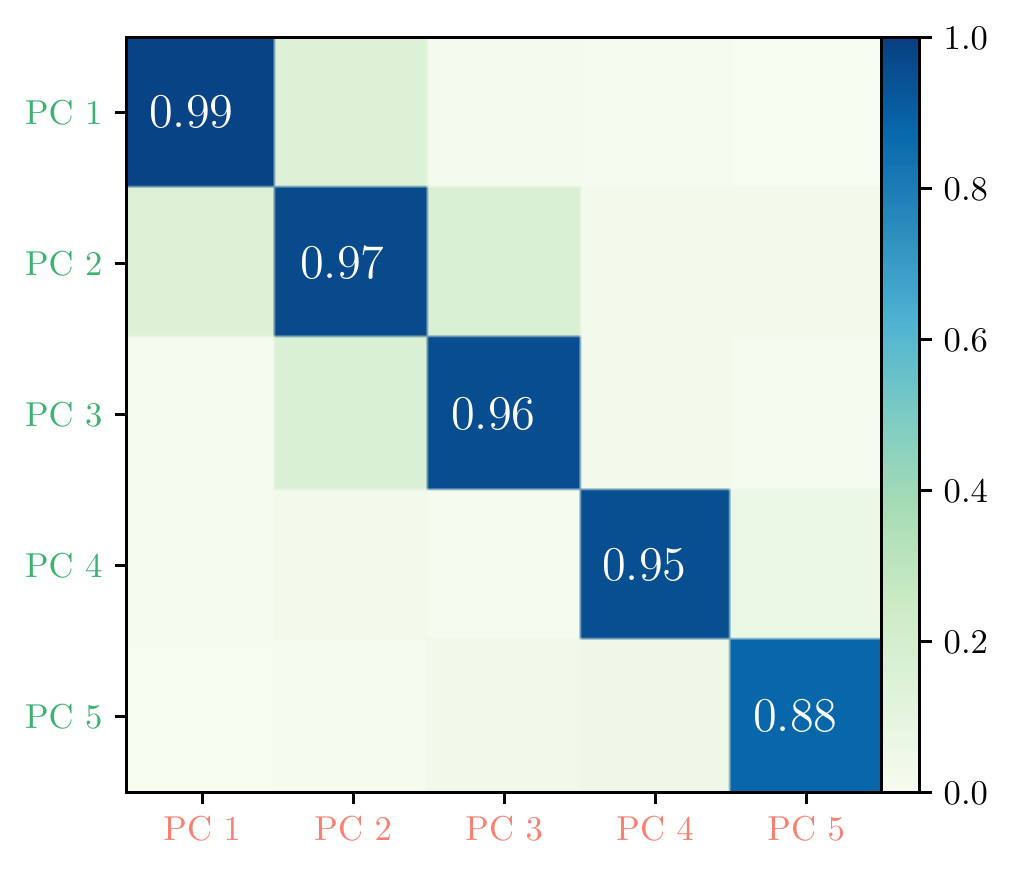}

    \caption{{Graphical representation of the mixing matrix M$_{ij}$ between the principal components of real and simulated spectra. }}
    \label{fig: pc_matrix}
\end{figure}

\subsubsection{Eigencoefficients}\label{coeff}

We can determine the relative contribution of each eigenspectrum to the observed spectrum by calculating the respective eigencoefficients, i.e., $ \mathrm{a}_i$ and $ \mathrm{b}_i$ from Equations~\ref{specdata} and~\ref{specsim}. Those are simply the scalar products of the eigenspectra with their normalized spectra. Figures~\ref{fig: coeff_sdss} and~\ref{fig: coeff_uspec} show the distributions of the five eigencoefficients. Figure~\ref{fig: coeff_sdss} shows the real spectra from SDSS (green) and the simulated \texttt{U\textsc{spec}} spectra (red) projected onto SDSS principal components (i.e., onto the basis set $\phi_i (\lambda)$ from Equation~\ref{specdata}). The coefficients distributions for both the real and the simulated spectra are overlapping. Also, all the coefficients are correlated to each other. However, the \texttt{U\textsc{spec}} eigencoefficients are occupying a larger parameter space in all five coefficients. This is even more evident when looking at Figure~\ref{fig: coeff_uspec}, where the spectra are projected onto \texttt{U\textsc{spec}} principal components (or the basis set $\psi_i (\lambda)$ from Equation~\ref{specsim}). The coefficients distributions still appear to be overlapping and correlations between different coefficients are visible. However, the regions occupied by \texttt{U\textsc{spec}} spectra are clearly larger than SDSS ones. This is due to the higher signal-to-noise of \texttt{U\textsc{spec}} simulated spectra, which can be also seen in Figure~\ref{fig: pc_comparison}. This signal-to-noise ratio brings the eigencoefficients to be sensitive to a wider variety in properties in the simulated galaxy population. This effect can be accounted for when matching the noise properties of the SDSS spectra. Furthermore, the differences in the eigencoefficients can also be used as distance measures to check the correctness of the inputs of the simulations. This is particularly evident when comparing the projected coefficients here described and those shown in Appendix~\ref{appendix}. In an upcoming publication, we will use the eigencoefficients defined here to better constrain the redshift distribution $n(z)$ presented in \citep{herbel2017}, in addition to the distance measures already outlined in \citep{herbel2017}.

\begin{figure}
\centering

	\includegraphics[width=130mm]{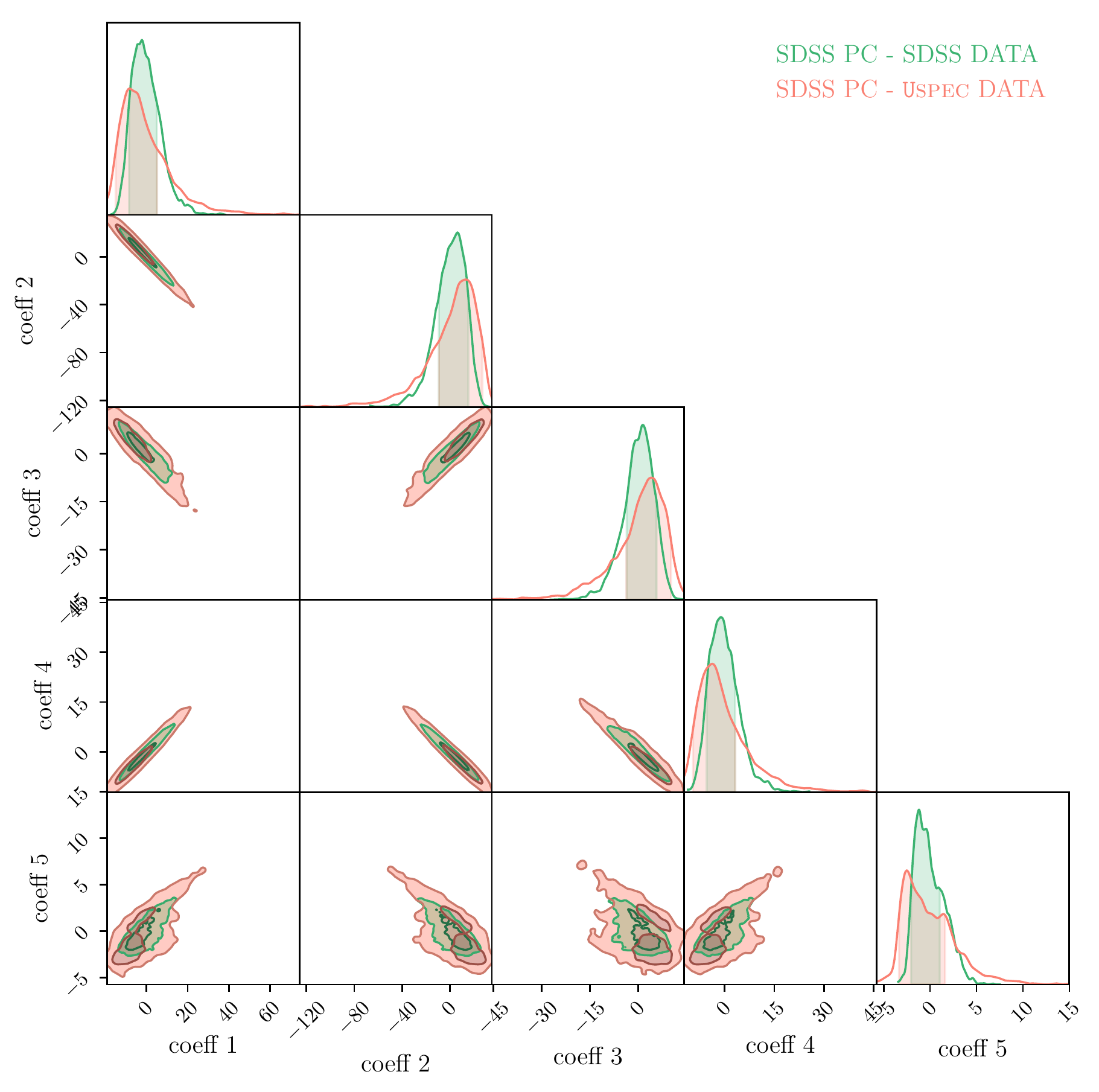}

    \caption{{Eigencoefficients resulting from projecting real spectra from SDSS (green) and simulated \texttt{U\textsc{spec}} spectra (red) onto SDSS principal components.}}
    \label{fig: coeff_sdss}
\end{figure}

\begin{figure}
\centering

	\includegraphics[width=130mm]{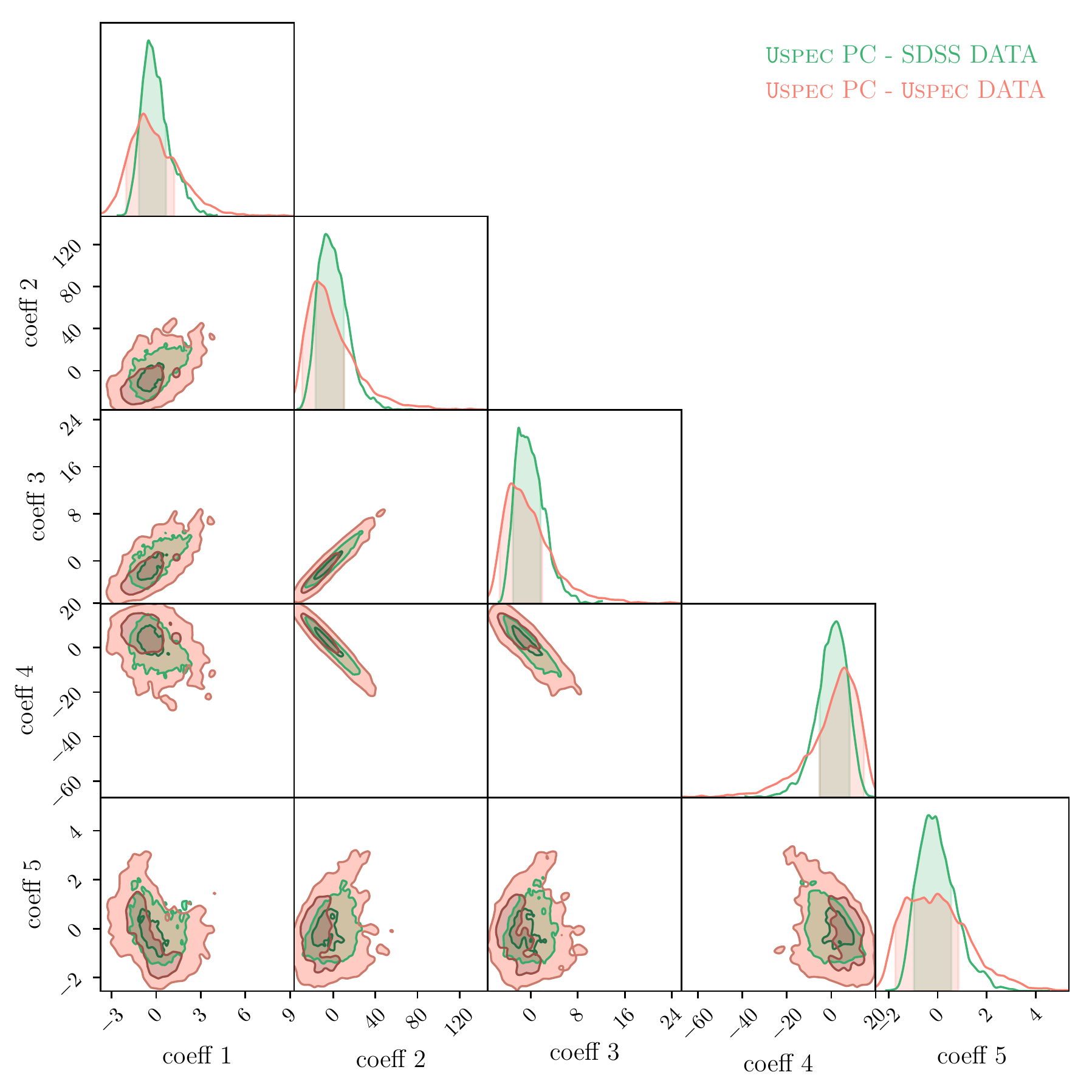}

    \caption{{Eigencoefficients resulting from projecting real spectra from SDSS (green) and simulated \texttt{U\textsc{spec}} spectra (red) onto \texttt{U\textsc{spec}} principal components. }}
    \label{fig: coeff_uspec}
\end{figure}

\section{Conclusions}\label{conclusions}

In this paper we describe \texttt{U\textsc{spec}}, a tool to simulate galaxy spectra for cosmological surveys. \texttt{U\textsc{spec}} builds galaxy spectra starting from a linear combination of $kcorrect$ templates and coefficients. The coefficients are chosen to match magnitudes and spectroscopic redshifts coming from luminosity functions which evolve with redshift. 

To compare our simulations to real data, we considered a LRG sample using the redshifts and colors cuts as in the CMASS sample from SDSS/BOSS. We apply these cuts to both real and simulated galaxies. We then modeled the noise and instrumental properties and included them in the simulated spectra. In particular, we modeled the read-out noise, the shot noise and the instrumental and atmospheric throughput. We also constructed and included in \texttt{U\textsc{spec}} a night background sky model spectrum with same characteristic as the ones observed at the APO in New Mexico. The LSF for the line broadening has been derived from sky spectra observed in the BOSS survey.

{We compared the average spectra of the real and simulated galaxy populations, finding  \texttt{U\textsc{spec}} able to reproduce realistic galaxy spectra of a given sample. The differences in spectral shape and emission lines visible in the two stacked spectra can be ascribed to the intrinsic input properties of the simulated galaxy population. Similar differences have also been reported in other attempts to study LRGs and in particular BOSS galaxies \citep{eisenstein2003, thomas2003, wake2006, maraston2009, tojeiro2012, maraston2013}. A forward modeling approach integrated into a full ABC (Approximate Bayesian Computation) framework and exploring possible impacts of more complex models of galaxy spectra which may be implemented in the run will help reducing these differences and will grant a better match of the overall galaxy populations. No attempt is done however to study  galaxy properties on an individual basis, as this goes beyond the scope of such simulations. The main purpose is to incorporate spectra simulations into a framework which includes broad-band and narrow-band data coming from different photometric and spectroscopic surveys, in order to better constrain the galaxy redshift distribution $n(z)$. $n(z)$ is of fundamental importance to improve our understanding of cosmological probes such as galaxy clustering and cosmic shear.}

To further quantify the level of agreement between the real and the simulated galaxy samples, we also performed a PCA. We find a remarkably good agreement between the two populations for {all the five principal components investigated here}. The comparison of the eigencoefficients of the two galaxy populations also shows that we are able to reproduce {at least part of} the variety of properties of LRGs in the SDSS survey. Both the distribution of the eigencoefficients resulting from projecting real spectra from SDSS and simulated \texttt{U\textsc{spec}} spectra onto SDSS, and onto \texttt{U\textsc{spec}} principal components, overlap. However, the \texttt{U\textsc{spec}} galaxies distributions are somewhat broader than those of data, which can be explained by the higher signal-to-noise ratio of the simulated galaxies, which results in a wider variety of projected coefficients. 

We define a distance measure $\varepsilon$ as the ratio between the product of the diagonal elements of the mixing matrix M$_{ij}$ of the PCAs and its determinant. The mixing matrix is expected to be the identity matrix if real and simulated data have the same principal components. We find $\varepsilon=0.95$. In the course of our analysis, we match the redshift distribution of real and simulated data. It is worth noting however that, if we have a difference in the median redshifts between the two galaxy sample of $\Delta z\sim0.018$, then the value of $\varepsilon$ decreases to 0.77. Also, the eigencoefficients distributions change when introducing such a difference. This is an interesting result as it indicates that the distance measure $\varepsilon$, as well as the eigencoefficients distributions, are sensitive to the parameters that control the redshift evolution of the input luminosity functions. For a detailed discussion on this aspect, see Appendix~\ref{appendix}.

The results presented here are promising and offer good prospects for applying our method to large upcoming spectroscopic surveys such as DESI. In our future work, we plan to incorporate \texttt{U\textsc{spec}} into a full ABC framework, to better match the intrinsic and noise properties of our simulated galaxies and real data. Furthermore, we will seek to simulate the population of blue star-forming galaxies, to also test the population of $\mathrm{[O\,II]}$ emitters which offers new different insights into the study of clustering of different galaxy populations. 

\acknowledgments
MF would like to thank Deniz Soyuer for carefully reading the manuscript. This research made use of \textsf{IP\textsc{ython}}, \textsf{\textsc{NumPy}},  \textsf{\textsc{SciPy}}, and \textsf{\textsc{Matplotlib}}. We acknowledge support by SNF grant $200021\_169130$.

\bibliographystyle{JHEP.bst}
\bibliography{/Users/martinafagioli/Documents/bib/bibliography.bib}

\providecommand{\href}[2]{#2}\begingroup\raggedright\begin{thebibliography}{10}

\bibitem{herbel2017}
J.~{Herbel}, T.~{Kacprzak}, A.~{Amara}, A.~{Refregier}, C.~{Bruderer} and
  A.~{Nicola}, \emph{{The redshift distribution of cosmological samples: a
  forward modeling approach}},
  \href{https://doi.org/10.1088/1475-7516/2017/08/035}{\emph{\jcap} {\bfseries
  8} (Aug., 2017) 035}, [\href{https://arxiv.org/abs/1705.05386}{{\ttfamily
  1705.05386}}].

\bibitem{peebles1970}
P.~J.~E. {Peebles} and J.~T. {Yu}, \emph{{Primeval Adiabatic Perturbation in an
  Expanding Universe}}, \href{https://doi.org/10.1086/150713}{\emph{\apj}
  {\bfseries 162} (Dec., 1970) 815}.

\bibitem{sunyaev1970}
R.~A. {Sunyaev} and Y.~B. {Zeldovich}, \emph{{Small scale entropy and adiabatic
  density perturbations {\mdash} Antimatter in the Universe}},
  \href{https://doi.org/10.1007/BF00649577}{\emph{\apss} {\bfseries 9} (Dec.,
  1970) 368--382}.

\bibitem{bond1984}
J.~R. {Bond} and G.~{Efstathiou}, \emph{{Cosmic background radiation
  anisotropies in universes dominated by nonbaryonic dark matter}},
  \href{https://doi.org/10.1086/184362}{\emph{\apjl} {\bfseries 285} (Oct.,
  1984) L45--L48}.

\bibitem{coil2013}
A.~L. {Coil}, \emph{{The Large-Scale Structure of the Universe}}, p.~387.
\newblock 2013.
\newblock 10.1007/978-94-007-5609-0\_8.

\bibitem{madgwick2003}
D.~S. {Madgwick}, E.~{Hawkins}, O.~{Lahav}, S.~{Maddox}, P.~{Norberg}, J.~A.
  {Peacock} et~al., \emph{{The 2dF Galaxy Redshift Survey: galaxy clustering
  per spectral type}},
  \href{https://doi.org/10.1046/j.1365-8711.2003.06861.x}{\emph{\mnras}
  {\bfseries 344} (Sept., 2003) 847--856},
  [\href{https://arxiv.org/abs/astro-ph/0303668}{{\ttfamily
  astro-ph/0303668}}].

\bibitem{zehavi2005}
I.~{Zehavi}, D.~J. {Eisenstein}, R.~C. {Nichol}, M.~R. {Blanton}, D.~W. {Hogg},
  J.~{Brinkmann} et~al., \emph{{The Intermediate-Scale Clustering of Luminous
  Red Galaxies}}, \href{https://doi.org/10.1086/427495}{\emph{\apj} {\bfseries
  621} (Mar., 2005) 22--31},
  [\href{https://arxiv.org/abs/astro-ph/0411557}{{\ttfamily
  astro-ph/0411557}}].

\bibitem{zehavi2011}
I.~{Zehavi}, Z.~{Zheng}, D.~H. {Weinberg}, M.~R. {Blanton}, N.~A. {Bahcall},
  A.~A. {Berlind} et~al., \emph{{Galaxy Clustering in the Completed SDSS
  Redshift Survey: The Dependence on Color and Luminosity}},
  \href{https://doi.org/10.1088/0004-637X/736/1/59}{\emph{\apj} {\bfseries 736}
  (July, 2011) 59}, [\href{https://arxiv.org/abs/1005.2413}{{\ttfamily
  1005.2413}}].

\bibitem{tegmark1997}
M.~{Tegmark}, \emph{{Measuring Cosmological Parameters with Galaxy Surveys}},
  \href{https://doi.org/10.1103/PhysRevLett.79.3806}{\emph{Physical Review
  Letters} {\bfseries 79} (Nov., 1997) 3806--3809},
  [\href{https://arxiv.org/abs/astro-ph/9706198}{{\ttfamily
  astro-ph/9706198}}].

\bibitem{goldberg1998}
D.~M. {Goldberg} and M.~A. {Strauss}, \emph{{Determination of the Baryon
  Density from Large-Scale Galaxy Redshift Surveys}},
  \href{https://doi.org/10.1086/305284}{\emph{\apj} {\bfseries 495} (Mar.,
  1998) 29--43}, [\href{https://arxiv.org/abs/astro-ph/9707209}{{\ttfamily
  astro-ph/9707209}}].

\bibitem{eisenstein1998}
D.~J. {Eisenstein}, W.~{Hu} and M.~{Tegmark}, \emph{{Cosmic Complementarity:
  H$_{0}$ and {$\Omega$}$_{m}$ from Combining Cosmic Microwave Background
  Experiments and Redshift Surveys}},
  \href{https://doi.org/10.1086/311582}{\emph{\apjl} {\bfseries 504} (Sept.,
  1998) L57--L60}, [\href{https://arxiv.org/abs/astro-ph/9805239}{{\ttfamily
  astro-ph/9805239}}].

\bibitem{hong2012}
T.~{Hong}, J.~L. {Han}, Z.~L. {Wen}, L.~{Sun} and H.~{Zhan}, \emph{{The
  Correlation Function of Galaxy Clusters and Detection of Baryon Acoustic
  Oscillations}}, \href{https://doi.org/10.1088/0004-637X/749/1/81}{\emph{\apj}
  {\bfseries 749} (Apr., 2012) 81},
  [\href{https://arxiv.org/abs/1202.0640}{{\ttfamily 1202.0640}}].

\bibitem{weiner2005}
B.~J. {Weiner}, A.~C. {Phillips}, S.~M. {Faber}, C.~N.~A. {Willmer}, N.~P.
  {Vogt}, L.~{Simard} et~al., \emph{{The DEEP Groth Strip Galaxy Redshift
  Survey. III. Redshift Catalog and Properties of Galaxies}},
  \href{https://doi.org/10.1086/427256}{\emph{\apj} {\bfseries 620} (Feb.,
  2005) 595--617}, [\href{https://arxiv.org/abs/astro-ph/0411128}{{\ttfamily
  astro-ph/0411128}}].

\bibitem{garilli2008}
B.~{Garilli}, O.~{Le F{\`e}vre}, L.~{Guzzo}, D.~{Maccagni}, V.~{Le Brun},
  S.~{de la Torre} et~al., \emph{{The Vimos VLT deep survey. Global properties
  of 20,000 galaxies in the I$_{AB}$ $\lt$ 22.5 WIDE survey}},
  \href{https://doi.org/10.1051/0004-6361:20078878}{\emph{\aap} {\bfseries 486}
  (Aug., 2008) 683--695}, [\href{https://arxiv.org/abs/0804.4568}{{\ttfamily
  0804.4568}}].

\bibitem{schlegel2009}
D.~{Schlegel}, M.~{White} and D.~{Eisenstein}, \emph{{The Baryon Oscillation
  Spectroscopic Survey: Precision measurement of the absolute cosmic distance
  scale}},  in \emph{astro2010: The Astronomy and Astrophysics Decadal Survey},
  vol.~2010 of \emph{ArXiv Astrophysics e-prints}, 2009,
  \href{https://arxiv.org/abs/0902.4680}{{\ttfamily 0902.4680}}.

\bibitem{eisenstein2011}
D.~J. {Eisenstein}, D.~H. {Weinberg}, E.~{Agol}, H.~{Aihara}, C.~{Allende
  Prieto}, S.~F. {Anderson} et~al., \emph{{SDSS-III: Massive Spectroscopic
  Surveys of the Distant Universe, the Milky Way, and Extra-Solar Planetary
  Systems}}, \href{https://doi.org/10.1088/0004-6256/142/3/72}{\emph{\aj}
  {\bfseries 142} (Sept., 2011) 72},
  [\href{https://arxiv.org/abs/1101.1529}{{\ttfamily 1101.1529}}].

\bibitem{dawson2016}
K.~S. {Dawson}, J.-P. {Kneib}, W.~J. {Percival}, S.~{Alam}, F.~D. {Albareti},
  S.~F. {Anderson} et~al., \emph{{The SDSS-IV Extended Baryon Oscillation
  Spectroscopic Survey: Overview and Early Data}},
  \href{https://doi.org/10.3847/0004-6256/151/2/44}{\emph{\aj} {\bfseries 151}
  (Feb., 2016) 44}, [\href{https://arxiv.org/abs/1508.04473}{{\ttfamily
  1508.04473}}].

\bibitem{zhao2016}
G.-B. {Zhao}, Y.~{Wang}, A.~J. {Ross}, S.~{Shandera}, W.~J. {Percival}, K.~S.
  {Dawson} et~al., \emph{{The extended Baryon Oscillation Spectroscopic Survey:
  a cosmological forecast}},
  \href{https://doi.org/10.1093/mnras/stw135}{\emph{\mnras} {\bfseries 457}
  (Apr., 2016) 2377--2390}, [\href{https://arxiv.org/abs/1510.08216}{{\ttfamily
  1510.08216}}].

\bibitem{desi1}
{DESI Collaboration}, A.~{Aghamousa}, J.~{Aguilar}, S.~{Ahlen}, S.~{Alam},
  L.~E. {Allen} et~al., \emph{{The DESI Experiment Part I: Science,Targeting,
  and Survey Design}}, {\emph{ArXiv e-prints} (Oct., 2016) },
  [\href{https://arxiv.org/abs/1611.00036}{{\ttfamily 1611.00036}}].

\bibitem{desi2}
{DESI Collaboration}, A.~{Aghamousa}, J.~{Aguilar}, S.~{Ahlen}, S.~{Alam},
  L.~E. {Allen} et~al., \emph{{The DESI Experiment Part II: Instrument
  Design}}, {\emph{ArXiv e-prints} (Oct., 2016) },
  [\href{https://arxiv.org/abs/1611.00037}{{\ttfamily 1611.00037}}].

\bibitem{laureijs2011}
R.~{Laureijs}, J.~{Amiaux}, S.~{Arduini}, J.~. {Augu{\`e}res}, J.~{Brinchmann},
  R.~{Cole} et~al., \emph{{Euclid Definition Study Report}}, {\emph{ArXiv
  e-prints} (Oct., 2011) }, [\href{https://arxiv.org/abs/1110.3193}{{\ttfamily
  1110.3193}}].

\bibitem{dejong2012}
R.~S. {de Jong}, O.~{Bellido-Tirado}, C.~{Chiappini}, {\'E}.~{Depagne},
  R.~{Haynes}, D.~{Johl} et~al., \emph{{4MOST: 4-metre multi-object
  spectroscopic telescope}},  in \emph{Ground-based and Airborne
  Instrumentation for Astronomy IV}, vol.~8446 of \emph{\procspie}, p.~84460T,
  Sept., 2012, \href{https://arxiv.org/abs/1206.6885}{{\ttfamily 1206.6885}},
  \href{https://doi.org/10.1117/12.926239}{DOI}.

\bibitem{campbell2004}
L.~{Campbell}, W.~{Saunders} and M.~{Colless}, \emph{{The tiling algorithm for
  the 6dF Galaxy Survey}},
  \href{https://doi.org/10.1111/j.1365-2966.2004.07745.x}{\emph{\mnras}
  {\bfseries 350} (June, 2004) 1467--1476},
  [\href{https://arxiv.org/abs/astro-ph/0403502}{{\ttfamily
  astro-ph/0403502}}].

\bibitem{blanton2003}
M.~R. {Blanton}, H.~{Lin}, R.~H. {Lupton}, F.~M. {Maley}, N.~{Young},
  I.~{Zehavi} et~al., \emph{{An Efficient Targeting Strategy for Multiobject
  Spectrograph Surveys: the Sloan Digital Sky Survey ``Tiling'' Algorithm}},
  \href{https://doi.org/10.1086/344761}{\emph{\aj} {\bfseries 125} (Apr., 2003)
  2276--2286}, [\href{https://arxiv.org/abs/astro-ph/0105535}{{\ttfamily
  astro-ph/0105535}}].

\bibitem{schlegel2011}
D.~{Schlegel}, F.~{Abdalla}, T.~{Abraham}, C.~{Ahn}, C.~{Allende Prieto},
  J.~{Annis} et~al., \emph{{The BigBOSS Experiment}}, {\emph{ArXiv e-prints}
  (June, 2011) }, [\href{https://arxiv.org/abs/1106.1706}{{\ttfamily
  1106.1706}}].

\bibitem{boller2012}
T.~{Boller} and T.~{Dwelly}, \emph{{The 4MOST facility simulator: instrument
  and science optimisation}},  in \emph{Observatory Operations: Strategies,
  Processes, and Systems IV}, vol.~8448 of \emph{\procspie}, p.~84480X, Sept.,
  2012, \href{https://arxiv.org/abs/1208.4733}{{\ttfamily 1208.4733}},
  \href{https://doi.org/10.1117/12.924818}{DOI}.

\bibitem{nord2016}
B.~{Nord}, A.~{Amara}, A.~{R{\'e}fr{\'e}gier}, L.~{Gamper}, L.~{Gamper},
  B.~{Hambrecht} et~al., \emph{{SPOKES: An end-to-end simulation facility for
  spectroscopic cosmological surveys}},
  \href{https://doi.org/10.1016/j.ascom.2016.02.001}{\emph{Astronomy and
  Computing} {\bfseries 15} (Apr., 2016) 1--15},
  [\href{https://arxiv.org/abs/1602.01480}{{\ttfamily 1602.01480}}].

\bibitem{refregier2014}
A.~{Refregier} and A.~{Amara}, \emph{{A way forward for Cosmic Shear:
  Monte-Carlo Control Loops}},
  \href{https://doi.org/10.1016/j.dark.2014.01.002}{\emph{Physics of the Dark
  Universe} {\bfseries 3} (Apr., 2014) 1--3},
  [\href{https://arxiv.org/abs/1303.4739}{{\ttfamily 1303.4739}}].

\bibitem{bruderer2017}
C.~{Bruderer}, A.~{Nicola}, A.~{Amara}, A.~{Refregier}, J.~{Herbel} and
  T.~{Kacprzak}, \emph{{Cosmic shear calibration with forward modeling}},
  {\emph{ArXiv e-prints} (July, 2017) },
  [\href{https://arxiv.org/abs/1707.06233}{{\ttfamily 1707.06233}}].

\bibitem{tortorelli2018}
L.~{Tortorelli}, L.~{Della Bruna}, J.~{Herbel}, A.~{Amara}, A.~{Refregier},
  A.~{Alarcon} et~al., \emph{{The PAU Survey: A Forward Modeling Approach for
  Narrow-band Imaging}}, {\emph{ArXiv e-prints} (May, 2018) },
  [\href{https://arxiv.org/abs/1805.05340}{{\ttfamily 1805.05340}}].

\bibitem{eisenstein2005}
D.~J. {Eisenstein}, I.~{Zehavi}, D.~W. {Hogg}, R.~{Scoccimarro}, M.~R.
  {Blanton}, R.~C. {Nichol} et~al., \emph{{Detection of the Baryon Acoustic
  Peak in the Large-Scale Correlation Function of SDSS Luminous Red Galaxies}},
  \href{https://doi.org/10.1086/466512}{\emph{\apj} {\bfseries 633} (Nov.,
  2005) 560--574}, [\href{https://arxiv.org/abs/astro-ph/0501171}{{\ttfamily
  astro-ph/0501171}}].

\bibitem{hutsi2006}
G.~{H{\"u}tsi}, \emph{{Power spectrum of the SDSS luminous red galaxies:
  constraints on cosmological parameters}},
  \href{https://doi.org/10.1051/0004-6361:20065377}{\emph{\aap} {\bfseries 459}
  (Nov., 2006) 375--389},
  [\href{https://arxiv.org/abs/astro-ph/0604129}{{\ttfamily
  astro-ph/0604129}}].

\bibitem{padmanabhan2007}
N.~{Padmanabhan}, D.~J. {Schlegel}, U.~{Seljak}, A.~{Makarov}, N.~A. {Bahcall},
  M.~R. {Blanton} et~al., \emph{{The clustering of luminous red galaxies in the
  Sloan Digital Sky Survey imaging data}},
  \href{https://doi.org/10.1111/j.1365-2966.2007.11593.x}{\emph{\mnras}
  {\bfseries 378} (July, 2007) 852--872},
  [\href{https://arxiv.org/abs/astro-ph/0605302}{{\ttfamily
  astro-ph/0605302}}].

\bibitem{almeida2008}
C.~{Almeida}, C.~M. {Baugh}, D.~A. {Wake}, C.~G. {Lacey}, A.~J. {Benson}, R.~G.
  {Bower} et~al., \emph{{Luminous red galaxies in hierarchical cosmologies}},
  \href{https://doi.org/10.1111/j.1365-2966.2008.13179.x}{\emph{\mnras}
  {\bfseries 386} (June, 2008) 2145--2160},
  [\href{https://arxiv.org/abs/0710.3557}{{\ttfamily 0710.3557}}].

\bibitem{norberg2002}
P.~{Norberg}, C.~M. {Baugh}, E.~{Hawkins}, S.~{Maddox}, D.~{Madgwick},
  O.~{Lahav} et~al., \emph{{The 2dF Galaxy Redshift Survey: the dependence of
  galaxy clustering on luminosity and spectral type}},
  \href{https://doi.org/10.1046/j.1365-8711.2002.05348.x}{\emph{\mnras}
  {\bfseries 332} (June, 2002) 827--838},
  [\href{https://arxiv.org/abs/astro-ph/0112043}{{\ttfamily
  astro-ph/0112043}}].

\bibitem{cappellari2004}
M.~{Cappellari} and E.~{Emsellem}, \emph{{Parametric Recovery of Line-of-Sight
  Velocity Distributions from Absorption-Line Spectra of Galaxies via Penalized
  Likelihood}}, \href{https://doi.org/10.1086/381875}{\emph{\pasp} {\bfseries
  116} (Feb., 2004) 138--147},
  [\href{https://arxiv.org/abs/astro-ph/0312201}{{\ttfamily
  astro-ph/0312201}}].

\bibitem{cappellari2017}
M.~{Cappellari}, \emph{{Improving the full spectrum fitting method: accurate
  convolution with Gauss-Hermite functions}},
  \href{https://doi.org/10.1093/mnras/stw3020}{\emph{\mnras} {\bfseries 466}
  (Apr., 2017) 798--811}, [\href{https://arxiv.org/abs/1607.08538}{{\ttfamily
  1607.08538}}].

\bibitem{eisenstein2003}
D.~J. {Eisenstein}, D.~W. {Hogg}, M.~{Fukugita}, O.~{Nakamura}, M.~{Bernardi},
  D.~P. {Finkbeiner} et~al., \emph{{Average Spectra of Massive Galaxies in the
  Sloan Digital Sky Survey}}, \href{https://doi.org/10.1086/346233}{\emph{\apj}
  {\bfseries 585} (Mar., 2003) 694--713},
  [\href{https://arxiv.org/abs/astro-ph/0212087}{{\ttfamily
  astro-ph/0212087}}].

\bibitem{thomas2003}
D.~{Thomas}, C.~{Maraston} and R.~{Bender}, \emph{{Stellar population models of
  Lick indices with variable element abundance ratios}},
  \href{https://doi.org/10.1046/j.1365-8711.2003.06248.x}{\emph{\mnras}
  {\bfseries 339} (Mar., 2003) 897--911},
  [\href{https://arxiv.org/abs/astro-ph/0209250}{{\ttfamily
  astro-ph/0209250}}].

\bibitem{lee2005}
H.-c. {Lee} and G.~{Worthey}, \emph{{{$\alpha$}-Enhanced Integrated Lick/IDS
  Spectral Indices and Milky Way and M31 Globular Clusters and Early-Type
  Galaxies}}, \href{https://doi.org/10.1086/432376}{\emph{\apjs} {\bfseries
  160} (Sept., 2005) 176--198},
  [\href{https://arxiv.org/abs/astro-ph/0504509}{{\ttfamily
  astro-ph/0504509}}].

\bibitem{thomas2005}
D.~{Thomas}, C.~{Maraston}, R.~{Bender} and C.~{Mendes de Oliveira}, \emph{{The
  Epochs of Early-Type Galaxy Formation as a Function of Environment}},
  \href{https://doi.org/10.1086/426932}{\emph{\apj} {\bfseries 621} (Mar.,
  2005) 673--694}, [\href{https://arxiv.org/abs/astro-ph/0410209}{{\ttfamily
  astro-ph/0410209}}].

\bibitem{thomas2010}
D.~{Thomas}, C.~{Maraston}, K.~{Schawinski}, M.~{Sarzi} and J.~{Silk},
  \emph{{Environment and self-regulation in galaxy formation}},
  \href{https://doi.org/10.1111/j.1365-2966.2010.16427.x}{\emph{\mnras}
  {\bfseries 404} (June, 2010) 1775--1789},
  [\href{https://arxiv.org/abs/0912.0259}{{\ttfamily 0912.0259}}].

\bibitem{onodera2015}
M.~{Onodera}, C.~M. {Carollo}, A.~{Renzini}, M.~{Cappellari}, C.~{Mancini},
  N.~{Arimoto} et~al., \emph{{The Ages, Metallicities, and Element Abundance
  Ratios of Massive Quenched Galaxies at z $\ge$ 1.6}},
  \href{https://doi.org/10.1088/0004-637X/808/2/161}{\emph{\apj} {\bfseries
  808} (Aug., 2015) 161}, [\href{https://arxiv.org/abs/1411.5023}{{\ttfamily
  1411.5023}}].

\bibitem{lonoce2015}
I.~{Lonoce}, M.~{Longhetti}, C.~{Maraston}, D.~{Thomas}, C.~{Mancini},
  A.~{Cimatti} et~al., \emph{{Old age and supersolar metallicity in a massive z
  $\sim$ 1.4 early-type galaxy from VLT/X-Shooter spectroscopy}},
  \href{https://doi.org/10.1093/mnras/stv2150}{\emph{\mnras} {\bfseries 454}
  (Dec., 2015) 3912--3919}, [\href{https://arxiv.org/abs/1509.04000}{{\ttfamily
  1509.04000}}].

\bibitem{fagioli2016}
M.~{Fagioli}, C.~M. {Carollo}, A.~{Renzini}, S.~J. {Lilly}, M.~{Onodera} and
  S.~{Tacchella}, \emph{{Minor Mergers or Progenitor Bias? The Stellar Ages of
  Small and Large Quenched Galaxies}},
  \href{https://doi.org/10.3847/0004-637X/831/2/173}{\emph{\apj} {\bfseries
  831} (Nov., 2016) 173}, [\href{https://arxiv.org/abs/1607.03493}{{\ttfamily
  1607.03493}}].

\bibitem{kriek2016}
M.~{Kriek}, C.~{Conroy}, P.~G. {van Dokkum}, A.~E. {Shapley}, J.~{Choi}, N.~A.
  {Reddy} et~al., \emph{{A massive, quiescent, population II galaxy at a
  redshift of 2.1}}, \href{https://doi.org/10.1038/nature20570}{\emph{\nat}
  {\bfseries 540} (Dec., 2016) 248--251},
  [\href{https://arxiv.org/abs/1612.02001}{{\ttfamily 1612.02001}}].

\bibitem{nicola2016}
A.~{Nicola}, A.~{Refregier} and A.~{Amara}, \emph{{Integrated approach to
  cosmology: Combining CMB, large-scale structure, and weak lensing}},
  \href{https://doi.org/10.1103/PhysRevD.94.083517}{\emph{\prd} {\bfseries 94}
  (Oct., 2016) 083517}, [\href{https://arxiv.org/abs/1607.01014}{{\ttfamily
  1607.01014}}].

\bibitem{nicola2017}
A.~{Nicola}, A.~{Refregier} and A.~{Amara}, \emph{{Integrated cosmological
  probes: Extended analysis}},
  \href{https://doi.org/10.1103/PhysRevD.95.083523}{\emph{\prd} {\bfseries 95}
  (Apr., 2017) 083523}, [\href{https://arxiv.org/abs/1612.03121}{{\ttfamily
  1612.03121}}].

\bibitem{johnston2011}
R.~{Johnston}, \emph{{Shedding light on the galaxy luminosity function}},
  \href{https://doi.org/10.1007/s00159-011-0041-9}{\emph{\aapr} {\bfseries 19}
  (Aug., 2011) 41}, [\href{https://arxiv.org/abs/1106.2039}{{\ttfamily
  1106.2039}}].

\bibitem{beare2015}
R.~{Beare}, M.~J.~I. {Brown}, K.~{Pimbblet}, F.~{Bian} and Y.-T. {Lin},
  \emph{{The $\lt$i$\gt$z$\lt$/i$\gt$ $\gt$ 1.2 Optical Luminosity Function
  from a Sample of 410,000 Galaxies in Bo{\#1255}tes}},
  \href{https://doi.org/10.1088/0004-637X/815/2/94}{\emph{\apj} {\bfseries 815}
  (Dec., 2015) 94}, [\href{https://arxiv.org/abs/1511.01580}{{\ttfamily
  1511.01580}}].

\bibitem{schechter1976}
P.~{Schechter}, \emph{{An analytic expression for the luminosity function for
  galaxies.}}, \href{https://doi.org/10.1086/154079}{\emph{\apj} {\bfseries
  203} (Jan., 1976) 297--306}.

\bibitem{blanton2005}
M.~R. {Blanton}, D.~J. {Schlegel}, M.~A. {Strauss}, J.~{Brinkmann},
  D.~{Finkbeiner}, M.~{Fukugita} et~al., \emph{{New York University Value-Added
  Galaxy Catalog: A Galaxy Catalog Based on New Public Surveys}},
  \href{https://doi.org/10.1086/429803}{\emph{\aj} {\bfseries 129} (June, 2005)
  2562--2578}, [\href{https://arxiv.org/abs/astro-ph/0410166}{{\ttfamily
  astro-ph/0410166}}].

\bibitem{dirichlet}
N.~Balakrishnan, \emph{Handbook of the Logistic Distribution}.
\newblock Statistics: A Series of Textbooks and Monographs. Taylor \& Francis,
  2013.

\bibitem{blanton2007}
M.~R. {Blanton} and S.~{Roweis}, \emph{{K-Corrections and Filter
  Transformations in the Ultraviolet, Optical, and Near-Infrared}},
  \href{https://doi.org/10.1086/510127}{\emph{\aj} {\bfseries 133} (Feb., 2007)
  734--754}, [\href{https://arxiv.org/abs/astro-ph/0606170}{{\ttfamily
  astro-ph/0606170}}].

\bibitem{bruzual2003}
G.~{Bruzual} and S.~{Charlot}, \emph{{Stellar population synthesis at the
  resolution of 2003}},
  \href{https://doi.org/10.1046/j.1365-8711.2003.06897.x}{\emph{\mnras}
  {\bfseries 344} (Oct., 2003) 1000--1028},
  [\href{https://arxiv.org/abs/astro-ph/0309134}{{\ttfamily
  astro-ph/0309134}}].

\bibitem{lee99}
D.~D. Lee and H.~S. Seung, \emph{Learning the parts of objects by nonnegative
  matrix factorization}, {\emph{Nature} {\bfseries 401} (1999) 788--791}.

\bibitem{chabrier2003}
G.~{Chabrier}, \emph{{Galactic Stellar and Substellar Initial Mass Function}},
  \href{https://doi.org/10.1086/376392}{\emph{\pasp} {\bfseries 115} (July,
  2003) 763--795}, [\href{https://arxiv.org/abs/astro-ph/0304382}{{\ttfamily
  astro-ph/0304382}}].

\bibitem{kewley2001}
L.~J. {Kewley}, M.~A. {Dopita}, R.~S. {Sutherland}, C.~A. {Heisler} and
  J.~{Trevena}, \emph{{Theoretical Modeling of Starburst Galaxies}},
  \href{https://doi.org/10.1086/321545}{\emph{\apj} {\bfseries 556} (July,
  2001) 121--140}, [\href{https://arxiv.org/abs/astro-ph/0106324}{{\ttfamily
  astro-ph/0106324}}].

\bibitem{berge2013}
J.~{Berg{\'e}}, L.~{Gamper}, A.~{R{\'e}fr{\'e}gier} and A.~{Amara}, \emph{{An
  Ultra Fast Image Generator (UFIG) for wide-field astronomy}},
  \href{https://doi.org/10.1016/j.ascom.2013.01.001}{\emph{Astronomy and
  Computing} {\bfseries 1} (Feb., 2013) 23--32},
  [\href{https://arxiv.org/abs/1209.1200}{{\ttfamily 1209.1200}}].

\bibitem{bruderer2016}
C.~{Bruderer}, C.~{Chang}, A.~{Refregier}, A.~{Amara}, J.~{Berg{\'e}} and
  L.~{Gamper}, \emph{{Calibrated Ultra Fast Image Simulations for the Dark
  Energy Survey}},
  \href{https://doi.org/10.3847/0004-637X/817/1/25}{\emph{\apj} {\bfseries 817}
  (Jan., 2016) 25}, [\href{https://arxiv.org/abs/1504.02778}{{\ttfamily
  1504.02778}}].

\bibitem{bonnett2016}
C.~{Bonnett}, M.~A. {Troxel}, W.~{Hartley}, A.~{Amara}, B.~{Leistedt}, M.~R.
  {Becker} et~al., \emph{{Redshift distributions of galaxies in the Dark Energy
  Survey Science Verification shear catalogue and implications for weak
  lensing}}, \href{https://doi.org/10.1103/PhysRevD.94.042005}{\emph{\prd}
  {\bfseries 94} (Aug., 2016) 042005},
  [\href{https://arxiv.org/abs/1507.05909}{{\ttfamily 1507.05909}}].

\bibitem{leistedt2016}
B.~{Leistedt}, H.~V. {Peiris}, F.~{Elsner}, A.~{Benoit-L{\'e}vy}, A.~{Amara},
  A.~H. {Bauer} et~al., \emph{{Mapping and Simulating Systematics due to
  Spatially Varying Observing Conditions in DES Science Verification Data}},
  \href{https://doi.org/10.3847/0067-0049/226/2/24}{\emph{\apjs} {\bfseries
  226} (Oct., 2016) 24}, [\href{https://arxiv.org/abs/1507.05647}{{\ttfamily
  1507.05647}}].

\bibitem{albareti2017}
F.~D. {Albareti}, C.~{Allende Prieto}, A.~{Almeida}, F.~{Anders},
  S.~{Anderson}, B.~H. {Andrews} et~al., \emph{{The 13th Data Release of the
  Sloan Digital Sky Survey: First Spectroscopic Data from the SDSS-IV Survey
  Mapping Nearby Galaxies at Apache Point Observatory}},
  \href{https://doi.org/10.3847/1538-4365/aa8992}{\emph{\apjs} {\bfseries 233}
  (Dec., 2017) 25}, [\href{https://arxiv.org/abs/1608.02013}{{\ttfamily
  1608.02013}}].

\bibitem{gunn2006}
J.~E. {Gunn}, W.~A. {Siegmund}, E.~J. {Mannery}, R.~E. {Owen}, C.~L. {Hull},
  R.~F. {Leger} et~al., \emph{{The 2.5 m Telescope of the Sloan Digital Sky
  Survey}}, \href{https://doi.org/10.1086/500975}{\emph{\aj} {\bfseries 131}
  (Apr., 2006) 2332--2359},
  [\href{https://arxiv.org/abs/astro-ph/0602326}{{\ttfamily
  astro-ph/0602326}}].

\bibitem{gunn1998}
J.~E. {Gunn}, M.~{Carr}, C.~{Rockosi}, M.~{Sekiguchi}, K.~{Berry}, B.~{Elms}
  et~al., \emph{{The Sloan Digital Sky Survey Photometric Camera}},
  \href{https://doi.org/10.1086/300645}{\emph{\aj} {\bfseries 116} (Dec., 1998)
  3040--3081}, [\href{https://arxiv.org/abs/astro-ph/9809085}{{\ttfamily
  astro-ph/9809085}}].

\bibitem{stoughton2002}
C.~{Stoughton}, R.~H. {Lupton}, M.~{Bernardi}, M.~R. {Blanton}, S.~{Burles},
  F.~J. {Castander} et~al., \emph{{Sloan Digital Sky Survey: Early Data
  Release}}, \href{https://doi.org/10.1086/324741}{\emph{\aj} {\bfseries 123}
  (Jan., 2002) 485--548}.

\bibitem{smee2013}
S.~A. {Smee}, J.~E. {Gunn}, A.~{Uomoto}, N.~{Roe}, D.~{Schlegel}, C.~M.
  {Rockosi} et~al., \emph{{The Multi-object, Fiber-fed Spectrographs for the
  Sloan Digital Sky Survey and the Baryon Oscillation Spectroscopic Survey}},
  \href{https://doi.org/10.1088/0004-6256/146/2/32}{\emph{\aj} {\bfseries 146}
  (Aug., 2013) 32}, [\href{https://arxiv.org/abs/1208.2233}{{\ttfamily
  1208.2233}}].

\bibitem{bolton2012}
A.~S. {Bolton}, D.~J. {Schlegel}, {\'E}.~{Aubourg}, S.~{Bailey}, V.~{Bhardwaj},
  J.~R. {Brownstein} et~al., \emph{{Spectral Classification and Redshift
  Measurement for the SDSS-III Baryon Oscillation Spectroscopic Survey}},
  \href{https://doi.org/10.1088/0004-6256/144/5/144}{\emph{\aj} {\bfseries 144}
  (Nov., 2012) 144}, [\href{https://arxiv.org/abs/1207.7326}{{\ttfamily
  1207.7326}}].

\bibitem{dawson2013}
K.~S. {Dawson}, D.~J. {Schlegel}, C.~P. {Ahn}, S.~F. {Anderson},
  {\'E}.~{Aubourg}, S.~{Bailey} et~al., \emph{{The Baryon Oscillation
  Spectroscopic Survey of SDSS-III}},
  \href{https://doi.org/10.1088/0004-6256/145/1/10}{\emph{\aj} {\bfseries 145}
  (Jan., 2013) 10}, [\href{https://arxiv.org/abs/1208.0022}{{\ttfamily
  1208.0022}}].

\bibitem{maraston2013}
C.~{Maraston}, J.~{Pforr}, B.~M. {Henriques}, D.~{Thomas}, D.~{Wake}, J.~R.
  {Brownstein} et~al., \emph{{Stellar masses of SDSS-III/BOSS galaxies at z
  $\sim$ 0.5 and constraints to galaxy formation models}},
  \href{https://doi.org/10.1093/mnras/stt1424}{\emph{\mnras} {\bfseries 435}
  (Nov., 2013) 2764--2792}, [\href{https://arxiv.org/abs/1207.6114}{{\ttfamily
  1207.6114}}].

\bibitem{maraston2009}
C.~{Maraston}, G.~{Str{\"o}mb{\"a}ck}, D.~{Thomas}, D.~A. {Wake} and R.~C.
  {Nichol}, \emph{{Modelling the colour evolution of luminous red galaxies -
  improvements with empirical stellar spectra}},
  \href{https://doi.org/10.1111/j.1745-3933.2009.00621.x}{\emph{\mnras}
  {\bfseries 394} (Mar., 2009) L107--L111},
  [\href{https://arxiv.org/abs/0809.1867}{{\ttfamily 0809.1867}}].

\bibitem{eisenstein2001}
D.~J. {Eisenstein}, J.~{Annis}, J.~E. {Gunn}, A.~S. {Szalay}, A.~J. {Connolly},
  R.~C. {Nichol} et~al., \emph{{Spectroscopic Target Selection for the Sloan
  Digital Sky Survey: The Luminous Red Galaxy Sample}},
  \href{https://doi.org/10.1086/323717}{\emph{\aj} {\bfseries 122} (Nov., 2001)
  2267--2280}, [\href{https://arxiv.org/abs/astro-ph/0108153}{{\ttfamily
  astro-ph/0108153}}].

\bibitem{Cannon2006}
R.~{Cannon}, M.~{Drinkwater}, A.~{Edge}, D.~{Eisenstein}, R.~{Nichol},
  P.~{Outram} et~al., \emph{{The 2dF-SDSS LRG and QSO (2SLAQ) Luminous Red
  Galaxy Survey}},
  \href{https://doi.org/10.1111/j.1365-2966.2006.10875.x}{\emph{\mnras}
  {\bfseries 372} (Oct., 2006) 425--442},
  [\href{https://arxiv.org/abs/astro-ph/0607631}{{\ttfamily
  astro-ph/0607631}}].

\bibitem{thomas2013}
D.~{Thomas}, O.~{Steele}, C.~{Maraston}, J.~{Johansson}, A.~{Beifiori},
  J.~{Pforr} et~al., \emph{{Stellar velocity dispersions and emission line
  properties of SDSS-III/BOSS galaxies}},
  \href{https://doi.org/10.1093/mnras/stt261}{\emph{\mnras} {\bfseries 431}
  (May, 2013) 1383--1397}, [\href{https://arxiv.org/abs/1207.6115}{{\ttfamily
  1207.6115}}].

\bibitem{reid2016}
B.~{Reid}, S.~{Ho}, N.~{Padmanabhan}, W.~J. {Percival}, J.~{Tinker},
  R.~{Tojeiro} et~al., \emph{{SDSS-III Baryon Oscillation Spectroscopic Survey
  Data Release 12: galaxy target selection and large-scale structure
  catalogues}}, \href{https://doi.org/10.1093/mnras/stv2382}{\emph{\mnras}
  {\bfseries 455} (Jan., 2016) 1553--1573},
  [\href{https://arxiv.org/abs/1509.06529}{{\ttfamily 1509.06529}}].

\bibitem{bertin1996}
E.~{Bertin} and S.~{Arnouts}, \emph{{SExtractor: Software for source
  extraction.}}, \href{https://doi.org/10.1051/aas:1996164}{\emph{\aaps}
  {\bfseries 117} (June, 1996) 393--404}.

\bibitem{schlegel1998}
D.~J. {Schlegel}, D.~P. {Finkbeiner} and M.~{Davis}, \emph{{Maps of Dust
  Infrared Emission for Use in Estimation of Reddening and Cosmic Microwave
  Background Radiation Foregrounds}},
  \href{https://doi.org/10.1086/305772}{\emph{\apj} {\bfseries 500} (June,
  1998) 525--553}, [\href{https://arxiv.org/abs/astro-ph/9710327}{{\ttfamily
  astro-ph/9710327}}].

\bibitem{tojeiro2012}
R.~{Tojeiro}, W.~J. {Percival}, D.~A. {Wake}, C.~{Maraston}, R.~A. {Skibba},
  I.~{Zehavi} et~al., \emph{{The progenitors of present-day massive red
  galaxies up to z $\approx$ 0.7 - finding passive galaxies using SDSS-I/II and
  SDSS-III}},
  \href{https://doi.org/10.1111/j.1365-2966.2012.21177.x}{\emph{\mnras}
  {\bfseries 424} (July, 2012) 136--156},
  [\href{https://arxiv.org/abs/1202.6241}{{\ttfamily 1202.6241}}].

\bibitem{wake2006}
D.~A. {Wake}, R.~C. {Nichol}, D.~J. {Eisenstein}, J.~{Loveday}, A.~C. {Edge},
  R.~{Cannon} et~al., \emph{{The 2df SDSS LRG and QSO survey: evolution of the
  luminosity function of luminous red galaxies to z = 0.6}},
  \href{https://doi.org/10.1111/j.1365-2966.2006.10831.x}{\emph{\mnras}
  {\bfseries 372} (Oct., 2006) 537--550},
  [\href{https://arxiv.org/abs/astro-ph/0607629}{{\ttfamily
  astro-ph/0607629}}].

\bibitem{appenzeller2009}
I.~{Appenzeller}, \emph{{High-Redshift Galaxies - Light from the Early
  Universe}}.
\newblock 2009,
  \href{https://doi.org/10.1007/978-3-540-75824-2}{10.1007/978-3-540-75824-2}.

\bibitem{law2016}
D.~R. {Law}, B.~{Cherinka}, R.~{Yan}, B.~H. {Andrews}, M.~A. {Bershady},
  D.~{Bizyaev} et~al., \emph{{The Data Reduction Pipeline for the SDSS-IV MaNGA
  IFU Galaxy Survey}},
  \href{https://doi.org/10.3847/0004-6256/152/4/83}{\emph{\aj} {\bfseries 152}
  (Oct., 2016) 83}, [\href{https://arxiv.org/abs/1607.08619}{{\ttfamily
  1607.08619}}].

\bibitem{hanuschik2003}
R.~W. {Hanuschik}, \emph{{A flux-calibrated, high-resolution atlas of optical
  sky emission from UVES}},
  \href{https://doi.org/10.1051/0004-6361:20030885}{\emph{\aap} {\bfseries 407}
  (Sept., 2003) 1157--1164}.

\bibitem{osterbrock1996}
D.~E. {Osterbrock}, J.~P. {Fulbright}, A.~R. {Martel}, M.~J. {Keane}, S.~C.
  {Trager} and G.~{Basri}, \emph{{Night-Sky High-Resolution Spectral Atlas of
  OH and O2 Emission Lines for Echelle Spectrograph Wavelength Calibration}},
  \href{https://doi.org/10.1086/133722}{\emph{\pasp} {\bfseries 108} (Mar.,
  1996) 277}.

\bibitem{noll2012}
S.~{Noll}, W.~{Kausch}, M.~{Barden}, A.~M. {Jones}, C.~{Szyszka},
  S.~{Kimeswenger} et~al., \emph{{An atmospheric radiation model for Cerro
  Paranal. I. The optical spectral range}},
  \href{https://doi.org/10.1051/0004-6361/201219040}{\emph{\aap} {\bfseries
  543} (July, 2012) A92}, [\href{https://arxiv.org/abs/1205.2003}{{\ttfamily
  1205.2003}}].

\bibitem{jones2013}
A.~{Jones}, S.~{Noll}, W.~{Kausch}, C.~{Szyszka} and S.~{Kimeswenger},
  \emph{{An advanced scattered moonlight model for Cerro Paranal}},
  \href{https://doi.org/10.1051/0004-6361/201322433}{\emph{\aap} {\bfseries
  560} (Dec., 2013) A91}, [\href{https://arxiv.org/abs/1310.7030}{{\ttfamily
  1310.7030}}].

\bibitem{onodera2012}
M.~{Onodera}, A.~{Renzini}, M.~{Carollo}, M.~{Cappellari}, C.~{Mancini},
  V.~{Strazzullo} et~al., \emph{{Deep Near-infrared Spectroscopy of Passively
  Evolving Galaxies at z $\gt$\~{} 1.4}},
  \href{https://doi.org/10.1088/0004-637X/755/1/26}{\emph{\apj} {\bfseries 755}
  (Aug., 2012) 26}, [\href{https://arxiv.org/abs/1206.1540}{{\ttfamily
  1206.1540}}].

\bibitem{odonnell1994}
J.~E. {O'Donnell}, \emph{{R$_{nu}$-dependent optical and near-ultraviolet
  extinction}}, \href{https://doi.org/10.1086/173713}{\emph{\apj} {\bfseries
  422} (Feb., 1994) 158--163}.

\bibitem{pcareview}
I.~T. Jolliffe and J.~Cadima, \emph{Principal component analysis: a review and
  recent developments},
  \href{https://doi.org/10.1098/rsta.2015.0202}{\emph{Philosophical
  Transactions of the Royal Society of London A: Mathematical, Physical and
  Engineering Sciences} {\bfseries 374} (2016) },
  [\href{https://arxiv.org/abs/http://rsta.royalsocietypublishing.org/content/374/2065/20150202.full.pdf}{{\ttfamily
  http://rsta.royalsocietypublishing.org/content/374/2065/20150202.full.pdf}}].

\bibitem{efstathiou1984}
G.~{Efstathiou} and S.~M. {Fall}, \emph{{Multivariate analysis of elliptical
  galaxies}}, \href{https://doi.org/10.1093/mnras/206.3.453}{\emph{\mnras}
  {\bfseries 206} (Jan., 1984) 453--464}.

\bibitem{murtagh1987}
F.~{Murtagh} and A.~{Heck}, eds., \emph{{Multivariate Data Analysis}}, vol.~131
  of \emph{Astrophysics and Space Science Library}, 1987.
\newblock 10.1007/978-94-009-3789-5.

\bibitem{yip2004}
C.~W. {Yip}, A.~J. {Connolly}, D.~E. {Vanden Berk}, Z.~{Ma}, J.~A. {Frieman},
  M.~{SubbaRao} et~al., \emph{{Spectral Classification of Quasars in the Sloan
  Digital Sky Survey: Eigenspectra, Redshift, and Luminosity Effects}},
  \href{https://doi.org/10.1086/425626}{\emph{\aj} {\bfseries 128} (Dec., 2004)
  2603--2630}, [\href{https://arxiv.org/abs/astro-ph/0408578}{{\ttfamily
  astro-ph/0408578}}].

\bibitem{chen2012}
Y.-M. {Chen}, G.~{Kauffmann}, C.~A. {Tremonti}, S.~{White}, T.~M. {Heckman},
  K.~{Kova{\v c}} et~al., \emph{{Evolution of the most massive galaxies to z=
  0.6 - I. A new method for physical parameter estimation}},
  \href{https://doi.org/10.1111/j.1365-2966.2011.20306.x}{\emph{\mnras}
  {\bfseries 421} (Mar., 2012) 314--332},
  [\href{https://arxiv.org/abs/1108.4719}{{\ttfamily 1108.4719}}].

\bibitem{connolly1995}
A.~J. {Connolly}, A.~S. {Szalay}, M.~A. {Bershady}, A.~L. {Kinney} and
  D.~{Calzetti}, \emph{{Spectral Classification of Galaxies: an Orthogonal
  Approach}}, \href{https://doi.org/10.1086/117587}{\emph{\aj} {\bfseries 110}
  (Sept., 1995) 1071},
  [\href{https://arxiv.org/abs/astro-ph/9411044}{{\ttfamily
  astro-ph/9411044}}].

\bibitem{trager1998}
S.~C. {Trager}, G.~{Worthey}, S.~M. {Faber}, D.~{Burstein} and J.~J.
  {Gonz{\'a}lez}, \emph{{Old Stellar Populations. VI. Absorption-Line Spectra
  of Galaxy Nuclei and Globular Clusters}},
  \href{https://doi.org/10.1086/313099}{\emph{\apjs} {\bfseries 116} (1998)
  1--28}, [\href{https://arxiv.org/abs/astro-ph/9712258}{{\ttfamily
  astro-ph/9712258}}].

\bibitem{burstein1984}
D.~{Burstein}, S.~M. {Faber}, C.~M. {Gaskell} and N.~{Krumm}, \emph{{Old
  stellar populations. I - A spectroscopic comparison of galactic globular
  clusters, M31 globular clusters, and elliptical galaxies}},
  \href{https://doi.org/10.1086/162718}{\emph{\apj} {\bfseries 287} (Dec.,
  1984) 586--609}.

\bibitem{worthey1994}
G.~{Worthey}, S.~M. {Faber}, J.~J. {Gonzalez} and D.~{Burstein}, \emph{{Old
  stellar populations. 5: Absorption feature indices for the complete LICK/IDS
  sample of stars}}, \href{https://doi.org/10.1086/192087}{\emph{\apjs}
  {\bfseries 94} (Oct., 1994) 687--722}.

\bibitem{thomasmaraston2003}
D.~{Thomas}, C.~{Maraston} and R.~{Bender}, \emph{{New clues on the calcium
  underabundance in early-type galaxies}},
  \href{https://doi.org/10.1046/j.1365-8711.2003.06659.x}{\emph{\mnras}
  {\bfseries 343} (July, 2003) 279--283},
  [\href{https://arxiv.org/abs/astro-ph/0303615}{{\ttfamily
  astro-ph/0303615}}].

\bibitem{thomas2011}
D.~{Thomas}, C.~{Maraston} and J.~{Johansson}, \emph{{Flux-calibrated stellar
  population models of Lick absorption-line indices with variable element
  abundance ratios}},
  \href{https://doi.org/10.1111/j.1365-2966.2010.18049.x}{\emph{\mnras}
  {\bfseries 412} (Apr., 2011) 2183--2198},
  [\href{https://arxiv.org/abs/1010.4569}{{\ttfamily 1010.4569}}].

\end{thebibliography}\endgroup

\appendix

\section{Sensitivity analysis for the redshift distribution}\label{appendix}

This paper is aimed at presenting \texttt{U\textsc{spec}} and its capabilities of simulating realistic galaxy spectra. The PCA analysis quantifies the differences between the real and simulated spectra. However, here we describe how the PCA analysis can be also used as an additional constraint for the input redshift distribution $n(z)$ of \citep{herbel2017}.

Figures~\ref{fig: pca_fullz} and~\ref{fig: coeff_fullz} show the PCA analysis results when keeping all the galaxies that pass the selection criteria described in Section~\ref{sampsel}, i.e., not matching the redshift distributions. This brings the total number of galaxies analyzed to 2680. The left panel of Figure~\ref{fig: pca_fullz} shows the comparison between the five principal components of real and simulated spectra. The first two components are only sensitive to the overall shape of the spectra, and appear to be almost unchanged with respect to those in Figure~\ref{fig: pc_comparison}. It is visible however how the position of the redshifted 4000 \AA{} break is changed for the \texttt{U\textsc{spec}} galaxies. This is especially evident in the first component. The effect of the redshift difference (as reported in the main text in Section~\ref{redshifts}, $ \mathrm{median}(z_{\mathrm{SDSS}})-\mathrm{median}(z_{\texttt{U\textsc{spec}}})=0.018$) is more evident in the higher order components, where the spectral features for the \texttt{U\textsc{spec}} spectra are shifted towards bluer wavelengths, where the median of the \texttt{U\textsc{spec}} $n(z)$ is centered. The difference is reflected in the mixing matrix M$_{ij}$, shown in the right panel of Figure~\ref{fig: pca_fullz}. If we evaluate the distance metrics based on the mixing matrix, we find:

\begin{equation}
\varepsilon=\norm {\frac{\Pi(\mathrm{M}_{ii})}{\det(\mathrm{M}_{ij})} }=0.77
\end{equation}
From the comparison with the evaluation of the mixing matrix in the main text ($\varepsilon=0.95$), where the $n(z)$ of real and simulated data are matched, it is clear how this can be used as a distance metrics to constrain the input $n(z)$.

Also, Figure~\ref{fig: coeff_fullz} shows the eigencoefficients both projected onto SDSS principal components (left panel), and onto \texttt{U\textsc{spec}} principal components (right panel). The distributions of higher order \texttt{U\textsc{spec}} coefficients are clearly different than those of SDSS, and also than those shown in the main text (Figures~\ref{fig: coeff_sdss} and \ref{fig: coeff_uspec}). This is an indication of how the coefficient distribution can be used as distance metrics in refining the input $n(z)$ of simulated galaxies.
		
\begin{figure}	
	\centering
	\begin{subfigure}[t]{0.53\linewidth}
		\centering
		\includegraphics[width=1\linewidth, valign=t]{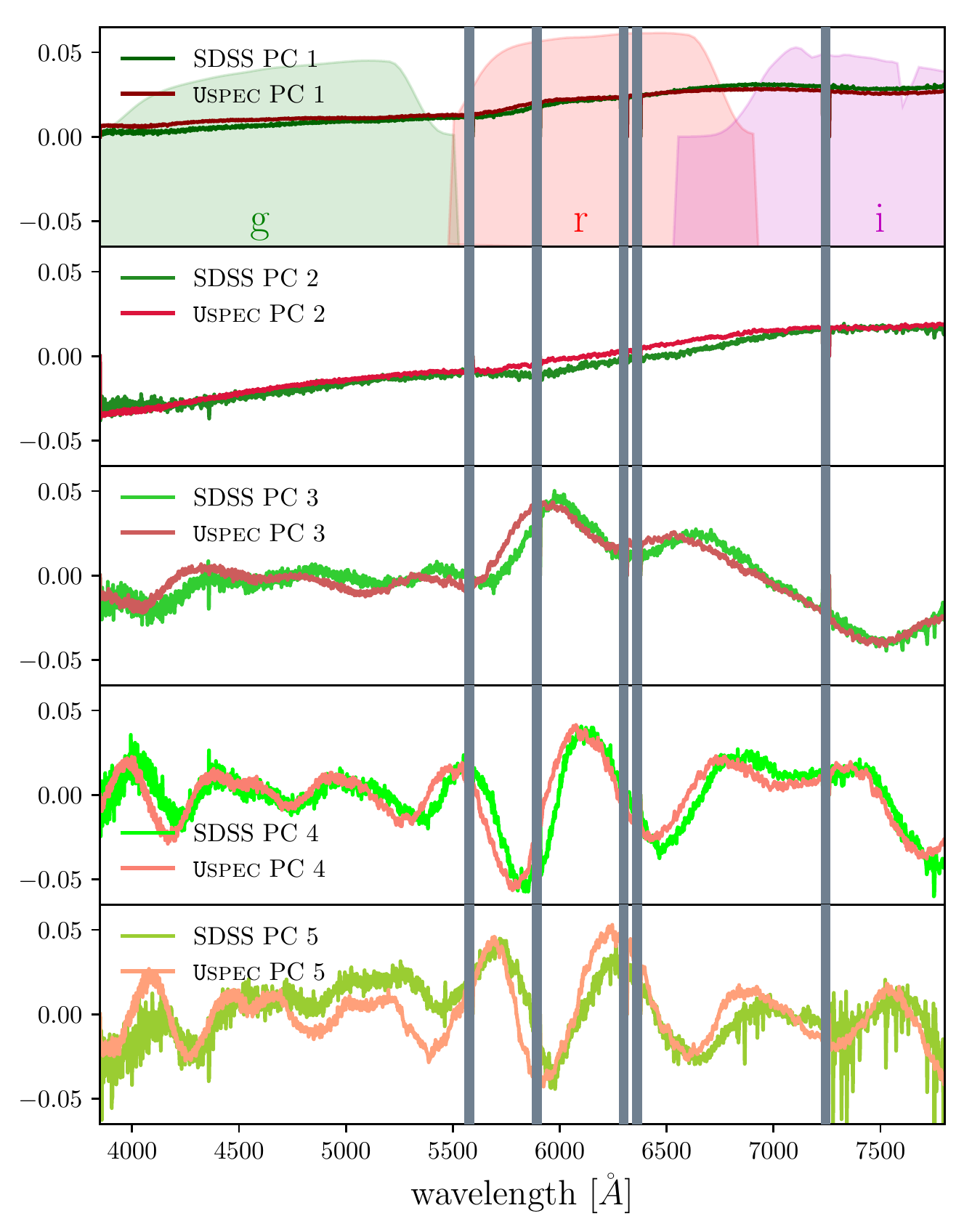}
	\end{subfigure}
	\begin{subfigure}[t]{0.43\linewidth}
		\centering
		\includegraphics[width=1\linewidth,valign=t]{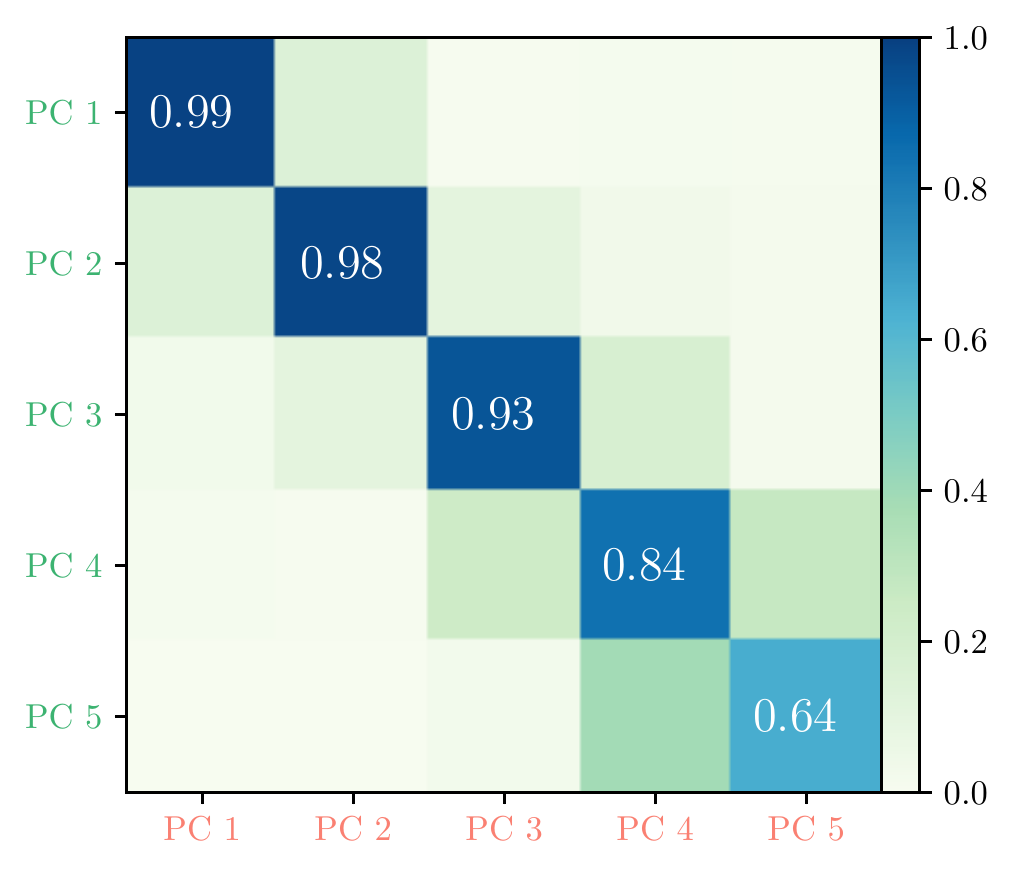}
		\caption{Mixing matrix for not-matched redshift distributions (i.e., keeping all the galaxies as shown in Figure~\ref{fig: redshifts}). Following formulas in Section~\ref{mixingmatrix}, we find:			
  		\begin{minipage}{\linewidth}
     		\begin{equation*}
        		\varepsilon= \norm {\frac{\Pi(\mathrm{M}_{ii})}{\det(\mathrm{M}_{ij})} }=0.77
     		\end{equation*}
 		 \end{minipage}
		}

	\label{fig:1b}

	\end{subfigure}
	\caption{{PCA comparison and mixing matrix for not-matched redshift distributions. The positions of broad features appear shifted when a difference in the median redshift distributions is introduced. This effect is especially evident in the last three principal components.}}\label{fig: pca_fullz}
\end{figure}

\begin{figure}	
	\centering
	\begin{subfigure}[t]{0.47\linewidth}
		\centering
		\includegraphics[width=1\linewidth, valign=t]{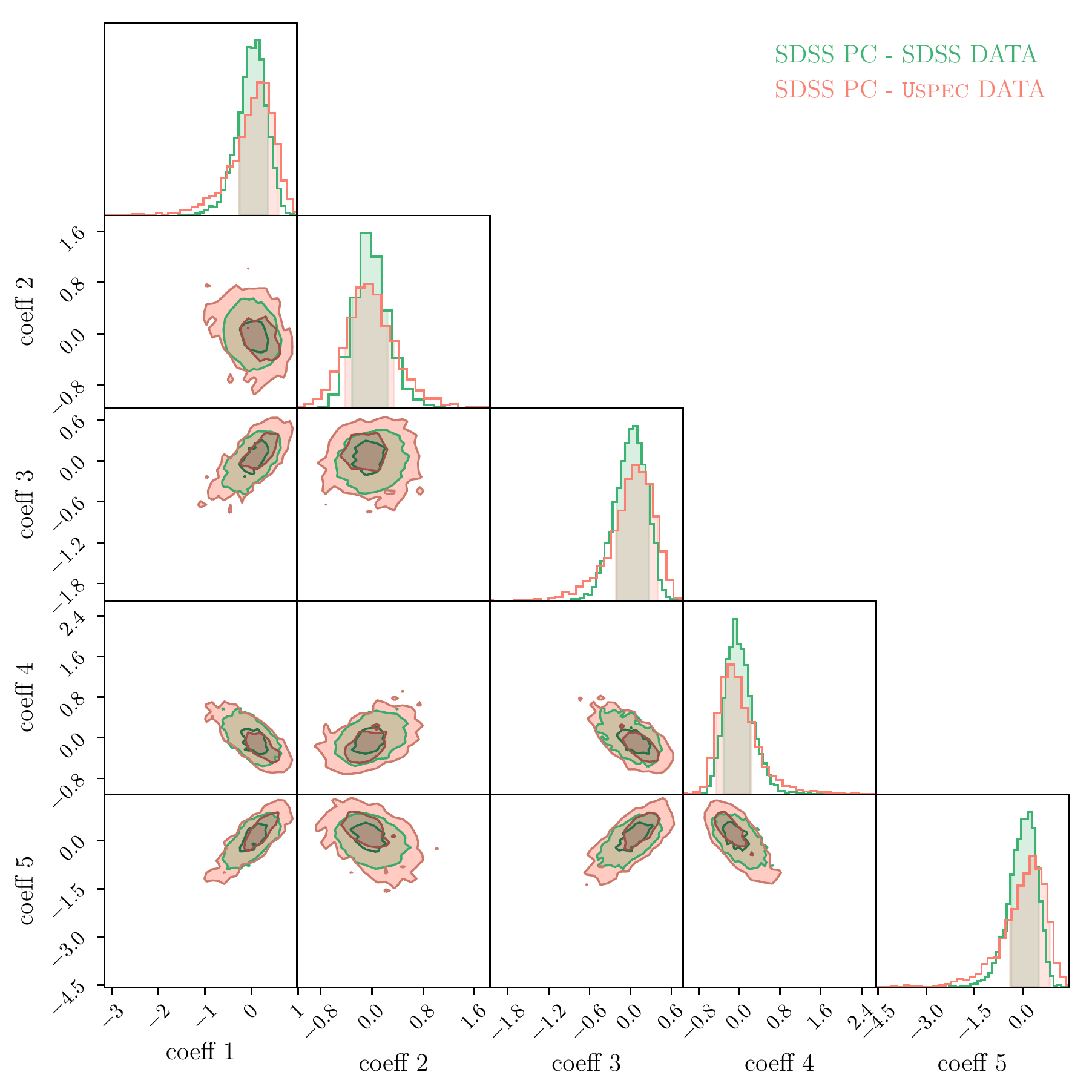}
	\end{subfigure}
	\begin{subfigure}[t]{0.47\linewidth}
		\centering
		\includegraphics[width=1\linewidth,valign=t]{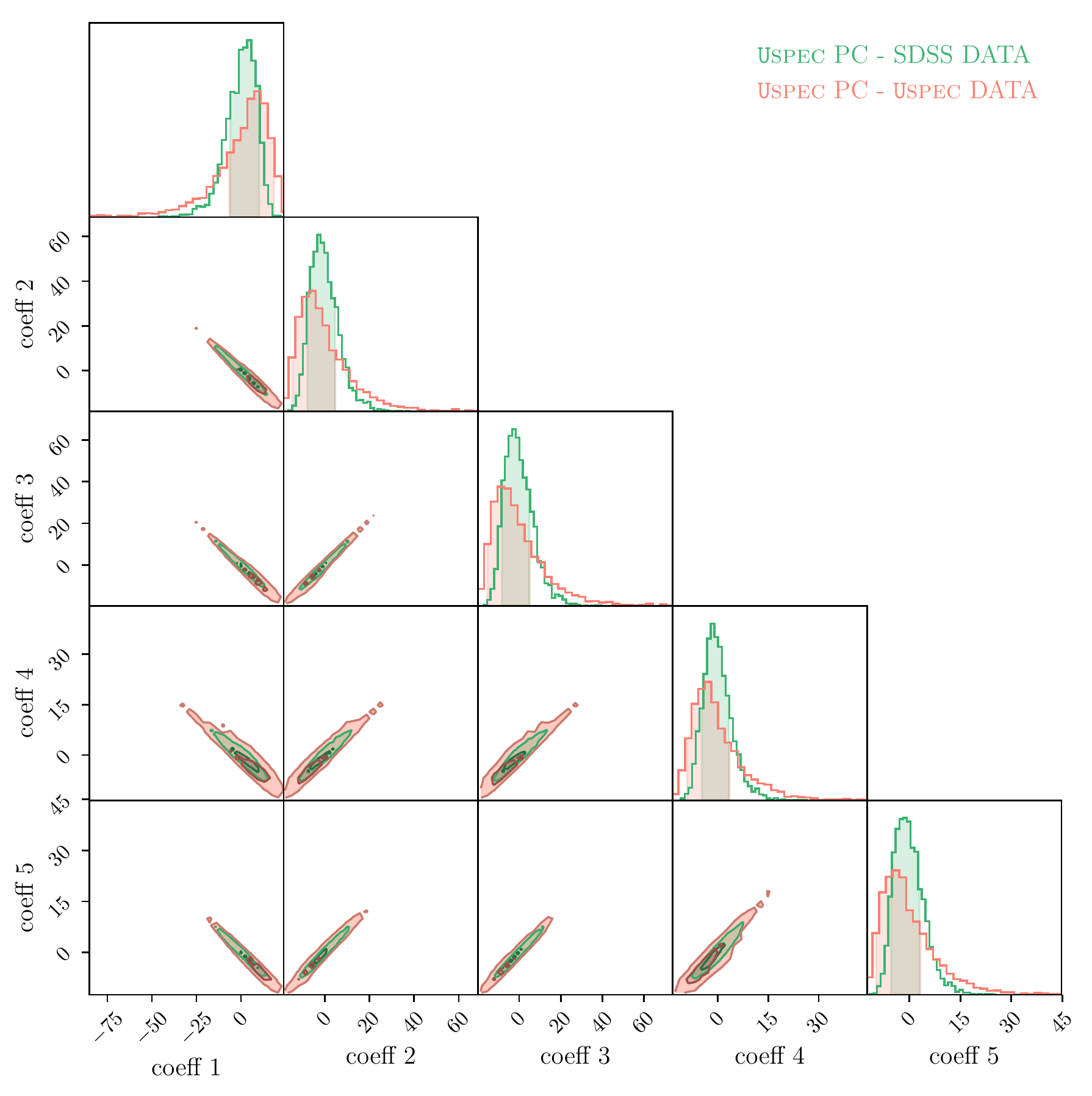}

	\end{subfigure}
	\caption{{Eigencoefficients projected onto SDSS principal components (left), and onto \texttt{U\textsc{spec}} principal components (right). The eigencoefficients can be used as distance measures to constrain the redshift distribution $n(z)$.}}
	\label{fig: coeff_fullz}
\end{figure}

\section{Impact of the sky model}\label{appendixb}

As discussed in Section~\ref{skymodel}, the background sky is an important source of systematics in any astronomical observation. The impossibility to predict the strength of sky lines can have important effects also on our PCA analysis. However, the position of the strong sky lines is easily predictable, as well as their width given the instrumental resolution R. Throughout our analysis, we masked the regions where the strongest sky lines at 5578.5 \AA{}, 5894.6 \AA{}, 6301.7 \AA{}, 6364.5 \AA{}, 7246 \AA{} are expected, with a FWHM of 15 \AA{} for each sky line. Figure~\ref{fig: pca_sky} shows how the PCA analysis looks line when not masking these sky lines. The redshift distributions are here matched. The first components appear mostly unchanged, excepted for the presence of the [O I] sky line at 5578.5 \AA{}. This sky line completely dominates the higher order principal components. This is reflected also in the mixing matrix shown in the right panel of Figure~\ref{fig: pca_sky}. In this case, $\varepsilon=0.001$. As sky lines are not redshifted, they are not washed out as galactic intrinsic emission lines when performing the PCA analysis in observed frame. This shows the impact of strong emission lines on the real spectra and the importance of properly masking them.

\begin{figure}	
	\centering
	\begin{subfigure}[t]{0.53\linewidth}
		\centering
		\includegraphics[width=1\linewidth, valign=t]{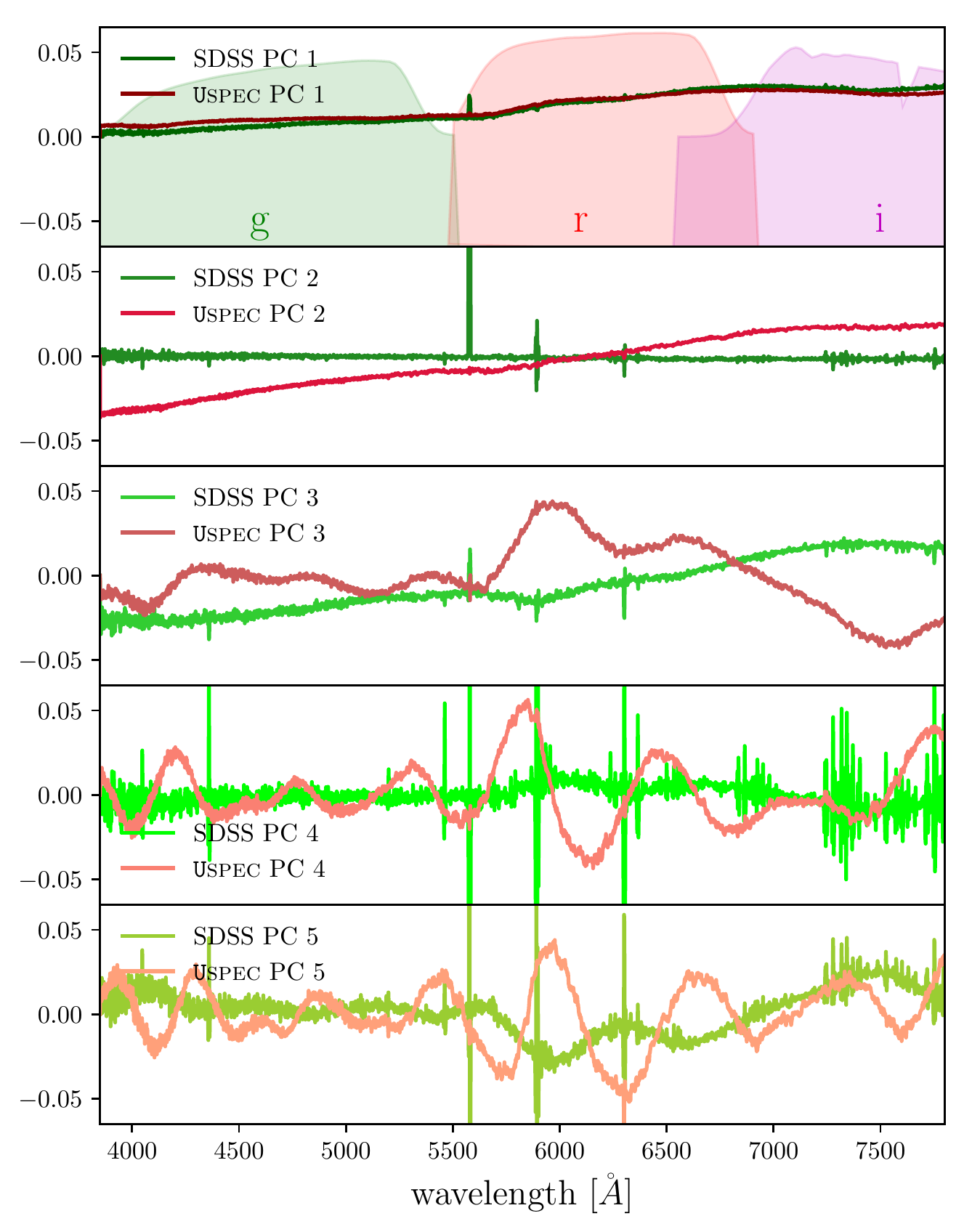}
	\end{subfigure}
	\begin{subfigure}[t]{0.43\linewidth}
		\centering
		\includegraphics[width=1\linewidth,valign=t]{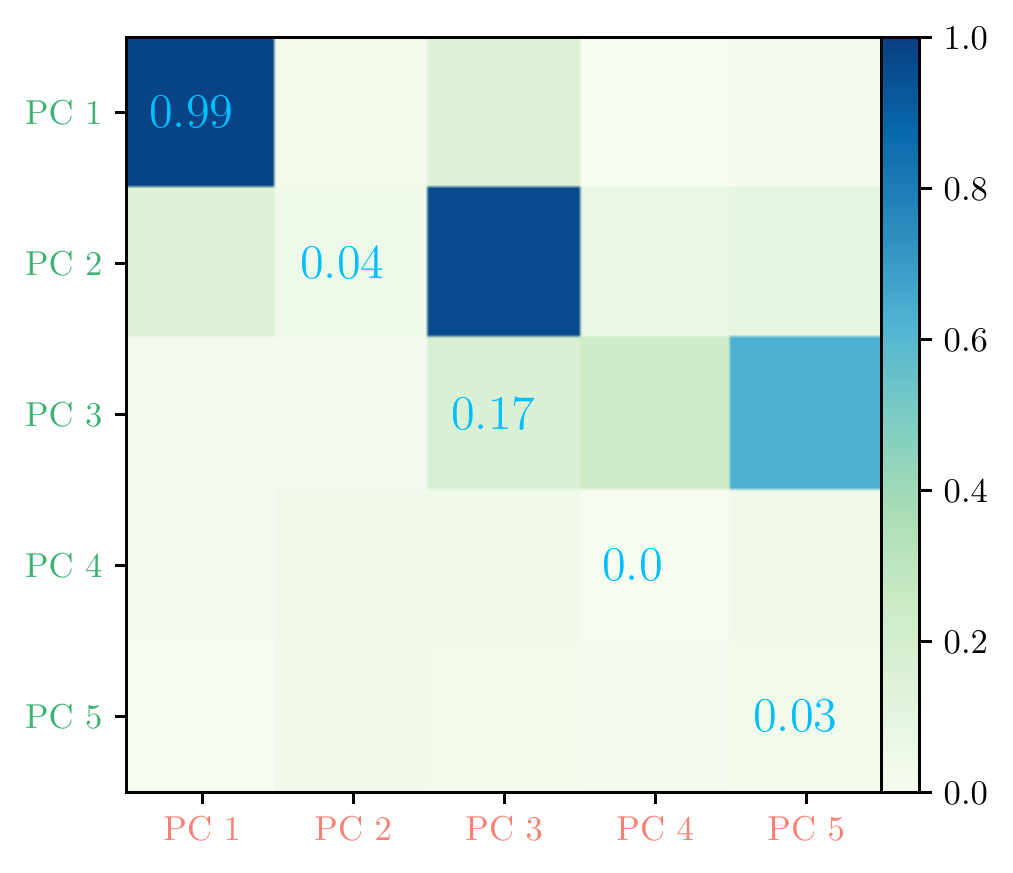}
		\caption{Mixing matrix when strong sky lines are not masked. The redshift distributions are here matched. Following formulas in Section~\ref{mixingmatrix}, we find:
  		\begin{minipage}{\linewidth}
     		\begin{equation*}
        		 \varepsilon=\norm {\frac{\Pi(\mathrm{M}_{ii})}{\det(\mathrm{M}_{ij})} }=0.001
     		\end{equation*}
 		 \end{minipage}
		}

	\label{fig:1b}

	\end{subfigure}
	\caption{{PCA comparison and mixing matrix when strong sky lines are not masked. While the first principal component is not sensitive to sky lines, the other principal components appear to be completely dominated by those for the SDSS population.}}
	\label{fig: pca_sky}
\end{figure}

\section{Stellar population properties of real and simulated galaxies}\label{appendixc}
To further explore the properties of our simulated galaxies, we computed some stellar population properties for our samples. In particular, we show here a comparison between the $<\mathrm{Fe}>$ (mean computed with indices Fe5270 and Fe5335) and Mg$b$ absorption line strengths for real and simulated galaxies. The absorption line strengths have been computed following the definition of pseudo-continua and bandpasses from \citep{trager1998} of the Lick System of spectral line indices (see among others \citep{burstein1984,worthey1994,thomasmaraston2003,thomas2005,thomas2003,thomas2011,fagioli2016}).  Figure~\ref{fig: fe_mgb} shows the broader variety of Fe and Mg abundances of real galaxies with respect to simulated ones; the simulated galaxy population falls within the parameter space of SDSS galaxies, being able to reproduce at least partially their [Mg$b$/Fe] abundance ratios. The mean of the Mg$b$ distribution of SDSS galaxies is shifted toward higher values with respect to the simulated galaxies, suggesting the need of a more complex modeling which includes super-solar abundance ratios in our framework.
\begin{figure}
\centering

	\includegraphics[width=90mm]{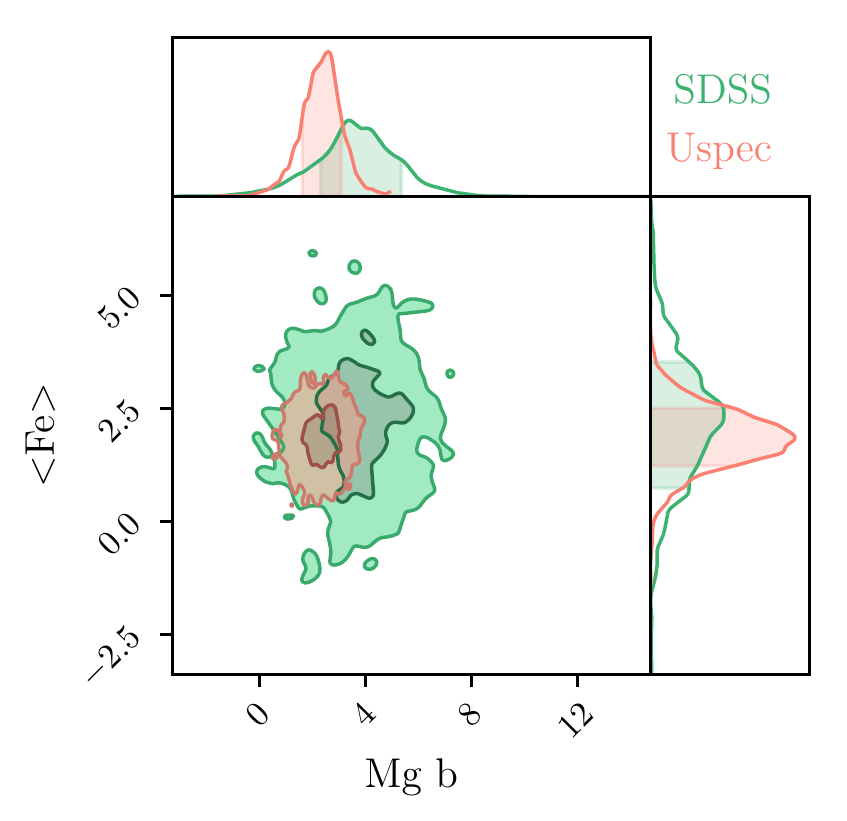}

    \caption{$<\mathrm{Fe}>$ vs Mg$b$ absorption line strengths for real (green) and simulated (red) galaxies.}
    \label{fig: fe_mgb}
\end{figure}

\end{document}